\documentclass[preprint]{elsarticle}
\journal{Elsevier}

\usepackage{graphicx}
\usepackage{latexsym}
\usepackage{color}
\usepackage{caption}
\usepackage{subcaption}
\usepackage{bm}
\usepackage{bbm}
\usepackage{amsmath,amssymb, amsthm}
\usepackage{algorithm}
\usepackage{algpseudocode}
\usepackage{color,soul}
\usepackage{graphics}
\usepackage{amsfonts}
\usepackage{setspace}   
\usepackage{wasysym}
\usepackage{multirow}

\usepackage{adjustbox}
\usepackage{graphicx}
\usepackage{multirow}
\usepackage{nomencl}
\usepackage{float}
\usepackage[table]{xcolor}% http://ctan.org/pkg/xcolor
\makenomenclature
\biboptions{sort&compress}
\newcommand*\diff{\mathop{}\!\mathrm{d}}

\usepackage{geometry}
\newtheorem{theorem}{Theorem}
\newtheorem{assumption}{Assumption}

\begin{document}

\begin{frontmatter}

\title{Divide and Conquer: An Incremental Sparsity Promoting Compressive Sampling Approach for Polynomial Chaos Expansions}

\author{Negin Alemazkoor}
\author{Hadi Meidani\corref{mycorrespondingauthor}}
\address{Department of Civil and Environmental Engineering, University of Illinois at Urbana-Champaign, Urbana, Illinois, USA.}
\cortext[mycorrespondingauthor]{Corresponding author}
\ead{meidani@illinois.edu}

\begin{abstract}
This paper introduces an efficient sparse recovery approach for Polynomial Chaos (PC)  expansions, which promotes the sparsity  by breaking the dimensionality of the problem. The proposed algorithm incrementally explores sub-dimensional expansions for a sparser recovery, and shows success when  removal of uninfluential parameters that results in a lower coherence for measurement matrix, allows for a higher order and/or sparser expansion to be recovered.  The incremental algorithm  effectively searches for the sparsest PC approximation, and not only can it decrease the prediction error, it can also reduce the dimensionality of PCE model. Four numerical examples are provided to demonstrate the validity of the proposed approach.  The results from  these examples show that the incremental algorithm substantially outperforms conventional compressive sampling approaches for PCE, in terms of both solution sparsity and prediction error.  \end{abstract}


\end{frontmatter}

%\linenumbers

%%\tableofcontents

\section{Introduction} \label{sec:introduction}

To reliably predict the behavior of a physical system of interest, it's vital to accurately model the interaction between the system's  inputs and outcomes, and account for the input uncertainty and its impact on the outcomes or the quantities of interests. The uncertain inputs to the model can include uncertain initial or boundary conditions, uncertain parameters, etc., which are often modeled as random variables. Uncertainty quantification (UQ) approaches are  used to model the propagation of random input variables through the model and quantify the uncertainty in the outputs. The most commonly known and used UQ approach is the Monte Carlo  approach, which despite its simple and robust applicability, suffers from a poor convergence rate of $\mathcal{O}(M^{-\frac{1}{2}})$, where $M$ is the number of samples. Although extensive research has been done and several modifications to the original Monte Carlo approach have been proposed to improve this convergence rate, e.g. importance sampling \cite{liu2008monte}, quasi-Monte Carlo \cite{caflisch1998monte}, multilevel Monte Carlo \cite{heinrich2001multilevel}) these sample-based approaches still suffer from slow rate of convergence. Therefore, there has been a growing interest in exploring alternative numerical approaches for UQ. 

Stochastic spectral approaches, as one alternative,  have been  shown to offer accelerated convergence and speed-up over Monte Carlo approaches, thereby being recognized as a promising solution to a variety of engineering problems especially when the dimensionality of  parameter space is not large. In these methods, the random output is represented through a spectral expansion on specific orthogonal basis functions such as generalized polynomial chaos (PC) \cite{ghanem2003stochastic, xiu2010numerical}. There are two distinguished classes of methods to evaluate the coefficients of the polynomial expansion, stochastic Galerkin and stochastic collocation methods. Stochastic Galerkin method is an extension of Galerkin approach, in which continuous operator problems such as differential equations are converted to a system of equations. Similar to Galerkin scheme, stochastic Galerkin leads to a $K$-coupled equation system, where $K$ is the number of coefficients to be estimated in polynomial chaos expansion (PCE), making it an intrusive approach. When the original problem has a complex nonlinear form, deriving the coupled equations can be cumbersome or even impossible.
 
Alternatively, in non-intrusive stochastic collocation methods, the coefficients are estimated either by solving a least square problem to fit a set of sample responses or by recasting the coefficients as integrals and approximating the integrals using sampling, quadrature, or sparse grid approaches. Despite the Galerkin method, the complexity of the original problem does not impact the applicability of collocation methods. However, due to the curse of dimensionality,  when the number of uncertain input parameters is large, even when efficient techniques such as sparse grid \cite{novak1997curse, barthelmann2000high, ganapathysubramanian2007sparse} are used, prohibitively large number of sample points are required for accurate estimation of chaos coefficients. Adaptive methods such as adaptive sparse grid \cite{hegland2003adaptive,gerstner2003dimension} and functional ANOVA decomposition \cite{foo2010multi,ma2010adaptive}  have been developed in order to reduce the required number of  samples by identifying the unimportant dimensions and placing fewer sample points in those dimensions.

Recently, additional efforts have been made to explore the regularity of the solution using a small number of samples when the solution is sparse and its approximated PCE has only few terms. Specifically, compressive sampling was used in \cite{doostan2011non,mathelin2012compressed,blatman2011adaptive} as a non-intrusive non-adapted approach to approximate the sparse solution of stochastic problems. Compressive sampling was first initiated in the field of signal and image reconstruction~\cite{candes2006robust,candes2006near,donoho2006compressed}. Conventionally, based on the distinguished Shannon's theorem, sufficient sampling rate was considered to be larger than twice the maximum frequency in the signal \cite{marks2012introduction}. Moreover, the fundamentals of linear algebra also impose the number of samples to be equal or larger than the dimensionality of the signal to ensure the reconstruction. However, compressive sampling techniques allowed reconstruction of sparse signals and images from incomplete measurements, i.e. a small set of samples, and effectively  solving an underdetermined linear system~\cite{candes2006robust,candes2006near,donoho2006compressed}.

In non-intrusive estimation of PCE, samples are acquired from an actual experimentation on the behavior of the system or from numerical simulations of a high-fidelity model for the system.  Both experimentation and simulation of complex systems can be very expensive, calling for efforts to reduce the number of required samples in  non-intrusive estimation procedures.  In \cite{doostan2011non,mathelin2012compressed}, compressive sampling was successfully applied in estimating the chaos coefficients  where the number of samples  was significantly smaller than the number of unknown chaos coefficients. However, this success heavily depends on two conditions: (a) the solution of stochastic problem should be in fact sufficiently sparse, and (b) the measurement matrix which includes random evaluations of polynomial bases should be sufficiently incoherent~\cite{doostan2011non}.  In \cite{yang2016enhancing}, the authors focused on improving the first condition, i.e.  the actual sparsity of the solution, by rotating the random inputs using the active subspace method~\cite{constantine2014active}.  As a result of this rotation, only a few influential  bases will effectively participate in the spectral representation of the quantity of interest (QoI), making the target representation sparser. However, as the authors admit, the coordinate rotation has negligible impact on the coherence of the measurement matrix. This means that the second condition for a successful sparse recovery is not improved. That is, no matter how effectively one can transform the coordinate system of the random inputs to enable a sparser \emph{target} solution, the attempts to recover that target solution using the available samples will still be in vain if the  measurement matrix is highly coherent. Our proposed approach tries to be aware of this second condition on the coherence. Specifically, instead of rotating the uncertainty sources, it concerns selecting for the polynomial expansion the right order and dimension, which are two factors that directly impact the coherence of the measurement matrix. This selection is done given a fixed sample set and basis type and an inaccuracy tolerance level. In what follows, we elaborate on how each one of these two factors can have implications on the success of  sparse recovery.
 
First, an increase in the total order for a PCE increases the number of columns in measurement matrix which makes it undesirably coherent. Therefore, the expansion order is usually chosen to be small, especially when the dimension of parameter space is large. This will deteriorate the accuracy if the exact solution is in fact of a higher order. In addition to the expansion order, an increase in the dimensionality also directly inflicts large coherence on measurement matrix, thereby degrading the accuracy of compressive sampling solution.  In \cite{jakeman2015enhancing}, an adaptive algorithm is introduced to identify and include only the important polynomial coordinates with the aim of  avoiding large coherence for the measurement matrix. Their algorithm begins with an initial basis set that includes all the dimensions  but have a low total order, typically equal to one. At each step of the algorithm, several anisotropic trial bases of higher orders in different dimensions are considered and the best one is selected to be included the basis set according to a cross-validation error criterion. 

In this work, we introduce a novel incremental compressive sampling algorithm, which considers the detrimental impact of coherence on solution's sparsity and accuracy, and searches for the sparsest solution by exploring through combinations of  dimensionality and order of a PCE. Compared to the adaptive algorithm proposed in~ \cite{jakeman2015enhancing}, our criterion for selecting new bases or stopping the algorithm is the sparsity of solutions calculated at each iterations. Also, the initial basis set in our algorithm can include as few as one dimension. This will remove the possibility of facing a large coherence hurdle at the initial step when high dimensional problems are concerned. This paper is organized as follows. Section~\ref{sec:CS in UQ} presents general concepts in compressive sampling and its theoretical background. In Section~\ref{sec:seq_cs}, we introduce our incremental compressive sampling algorithm along with relevant theoretical supports. Finally, Section~\ref{sec:numericalresults} includes numerical examples and discussions. 

\section{Compressive sampling in spectral uncertainty analysis} \label{sec:CS in UQ}

\subsection{Spectral uncertainty analysis} \label{sec:spectral methods}
Spectral uncertainty analysis methods aim to efficiently propagate the input uncertainty in numerical models. These methods are constituted based on expansions on orthogonal polynomial bases and have been proved to have spectral convergence \cite{ghanem2003stochastic,xiu2010numerical,babuska2004galerkin,xiu2002wiener,deb2001solution}. Formally, let $(\Omega,\mathcal{F},\mathbb{P})$ where $\Omega$ is the event space, $\mathcal{F}$ is a $\sigma$-field over $\Omega$, and $\mathbb{P}$ is the probability measure on $(\Omega,\mathcal{F})$. Also, let $\boldsymbol{\Xi}=(\Xi_{1}, ..., \Xi_{d})$ be a $d$-dimensional random vector with mutually independent components, joint density function,  $f_{\bm \Xi}(\bm \xi)$, or in short $\rho(\bm \xi)$, and finite moments on $\Omega$. Then any square-integrable random quantity of interest $u(\boldsymbol{\Xi}): \Omega \rightarrow \mathbb{R}$ can be represented as: 
\begin{equation}\label{eq:repUasSum}
u(\bm{\Xi})= \sum_{\bm \alpha \in \mathbb{N}_0^d} c_{\bm \alpha } \psi_{\bm \alpha}(\bm\Xi), 
\end{equation}
where   $\{ \psi_{\bm \alpha} \}_{\bm \alpha \in \mathbb{N}_0^d}$ is the set of orthogonal basis functions satisfying

\begin{equation} \label{eq:orogonality}
\int_{\Omega} \psi_{\bm n}(\bm \xi) \psi_{\bm m}(\bm \xi)  \rho({\bm \xi}) \diff \bm \xi =\gamma_{\bm n} \delta_{\bm {mn}}, \quad \bm{m,n} \in \mathbb{N}_0^d, 
\end{equation}
where $\gamma_{\bm n}$ is the normalization factor and $\delta_{\bm {mn}}$ is the delta function.

For computational sake, finite order truncation of spectral expansion is typically used to approximate $u(\bm \Xi)$:
\begin{equation}\label{eq:PCEexpansion}
u_{k}(\bm\Xi) := \sum_{\bm \alpha \in \Lambda_{d,k}} c_{\bm \alpha } \psi_{\bm \alpha}(\bm \Xi),
\end{equation}
where $\Lambda_{d,k}$ is the set of multi-indices defined as
\begin{equation} \label{eq:basisSet}
\Lambda_{d,k} := \{\bm \alpha \in \mathbb{N}_0^d : \Vert \bm \alpha \Vert_1 \le k \}.
\end{equation}

The cardinality of $\Lambda_{d,p}$, i.e., the  number of expansion terms, here denoted by $K$, is a function of $d$ and $k$ according to
\begin{equation}\label{eq:NumOfbasis}
K:= \vert \Lambda_{d,k} \vert = \frac{(k+d)!}{k!d!}.
\end{equation}
Given this setting, $u_{k}$ approximates $u$ in a proper sense and is referred to as the $k$-th degree PC approximation of $u$.

Each of the $K$ coefficients involved in $u_{k}$ definition can be exactly calculated by projecting $u$ onto the associated basis function:

\begin{equation}\label{eq:exactCoeff}
c_{\bm \alpha} = \frac{1}{\gamma_{\bm \alpha}}\int_{\Omega}u(\bm \xi)\psi_{\bm \alpha}(\bm \xi) \rho(\bm \xi) \diff \bm \xi.
\end{equation}
The exact calculation of above integral is a non-trivial and usually impossible task. Alternatively, the integral can be approximated using sampling, cubature rule or sparse grid techniques. However, when $d$ is large, even in efficient approaches such as sparse grid \cite{novak1997curse, barthelmann2000high, ganapathysubramanian2007sparse, constantine2012sparse}, the required number of samples is large. For instance, in Clenshaw-Curtis sparse grid, the number of samples need to be about $2^{k}$ times the number of expansion coefficients  \cite{barthelmann2000high}. Acquiring such a large sample set can be infeasible, due to limitations in experiments for empirical samples or limitations in computational resources for the case where samples are obtained from high-fidelity simulations. 

As briefly mentioned in Section~\ref{sec:introduction}, linear regression can also be used to approximate the expansion coefficients. Regression requires the number of collocation points to be at least equal to $K$, the number of expansion terms. Although  regression techniques significantly alleviate the prohibitive requirement on the number of samples, there has been concerns regarding its accuracy~\cite{xiu2010numerical}. Therefore, it is suggested to use at least $2K$ collocation points \cite{hosder2007efficient}. However, recently it has been shown that when the solution expansion is sparse, not only regression leads to accurate approximations but also the coefficients can be accurately estimated using samples much fewer than the number of coefficients, thereby saving costs in experimentation or high-fidelity simulation. This section follows with the review on the basics of sparse recovery for polynomial chaos expansions. 

\subsection{Sparse spectral analysis using compressive sampling} \label{sec:sparse recovery}
Compressive sampling has recently emerged and witnessed growing interest as an efficient and inexpensive signal processing algorithm, particularly applicable in cases where the issue of limited number of samples is present because of slow or expensive sampling procedures \cite{lustig2007sparse, paredes2007ultra, ender2010compressive, berger2010application, gemmeke2010compressive}. When the signal of interest is sparse, compressive sampling allows signal reconstruction with substantially fewer samples than the Shannon sampling rate, thereby suggesting promising solution for signal recovery when there are a few number of measurements available \cite{baraniuk2007compressive, candes2006compressive, cande2008introduction}. 

In the case of spectral stochastic techniques, we seek to fit an analytical model to a sample set obtained from the simulation of a complex high-fidelity model, or from expensive experiments. As a result, the generation of samples is a costly procedure, making compressive sampling an applicable algorithm to these problems. Due to its past successful records in reducing the number of required samples, compressive sampling continues to gain extensive attention in UQ. Particularly, it has been shown that if $u(\bm \Xi)$ is sparse with respect to polynomial bases,  the approximation expansion can be constructed using $ M\ll K$  random realizations \cite{doostan2011non, mathelin2012compressed,hampton2015compressive, yan2012stochastic, yang2013reweighted, peng2014weighted}. 

Let us denote random realizations of $\bm \Xi$ by $\{\bm \xi^{(i)}\}$. Also, let $\{u(\bm \xi^{(i)})\}$ and $\{u_{k}(\bm \xi^{(i)})\}$ respectively denote the  evaluations of $u(\bm \Xi)$ and $u_{k}(\bm \Xi)$ at those realizations. The sample vector $\{u(\bm \xi^{(i)})\}$ can be thought of as the set of empirical samples or the set of simulation results from an exact of high-fidelity model. We define vectors $\bm u$ and $\bm u_{k}$ and matrix $\bm \Psi$ as following:
\begin{equation} \label{eq:uVector}
\bm u:=(u(\bm \xi^{(1)}), ..., u(\bm \xi^{(M)}))^{T},
 \end{equation}
\begin{equation} \label{eq:ukVector}
\bm u_{k}:=(u_{k}(\bm \xi^{(1)}), ..., u_{k}(\bm \xi^{(M)}))^{T},
\end{equation}
\begin{equation} \label{eq:measurementMatrix}
\bm \Psi(i,j):= \psi_{\bm \alpha^{j}}(\bm \xi^{(i)}), \quad i=1,\cdots, M,\quad j=1:\cdots, K.
\end{equation}

Matrix $\bm \Psi$ has been referred to as measurement matrix or dictionary in the literature. The above definitions along with Equation~\ref{eq:PCEexpansion} simply imply that $\bm u_{k}=\bm \Psi \bm c$, where $\bm c \in \mathbb{R^{K}}$ is the vector of PC coefficients $c_{\bm \alpha}$. In order for $u_{k}(\bm \Xi)$ to approximate $u(\bm \Xi)$, the vector $\bm c$ should satisfy the condition $\left \| \bm u - \bm\Psi\bm c \right \|_{2}\leqslant  \epsilon$, where $\epsilon$ is the accuracy tolerance due to truncation error. Since we are interested in the case where $M \ll K$, there exist infinitely many solutions for $\bm c$. When $\bm c$ is sparse and has few significant terms, compressive sampling approach can be borrowed to find the sparsest solution. The sparse recovery approach can be formulated as the following program
\begin{equation} \label{eq:l0min}
\underset{\bm c}{\textrm{min}}\left \| \bm c \right \|_{0} \quad \textrm{subject to} \quad \left \| \bm u-\bm \Psi \bm c \right \|_{2}\leqslant \epsilon,
\end{equation}
where $\left \| \cdot \right \|_{0}$ denotes the $\ell_{0}$-norm, i.e. the number of non-zero components in a vector. This  $\ell_{0}$ minimization problem  is NP-hard to solve. Therefore, the non-covex discontinuous $\ell_{0}$-norm is usually replaced with convex continuous $\ell_{1}$-norm, leading to the  linear program
\begin{equation} \label{eq:consl1min}
\underset{\bm c}{\textrm{min}}\left \| \bm c \right \|_{1} \quad \textrm{subject to} \quad \left \| \bm u-\bm \Psi \bm c \right \|_{2}\leqslant \epsilon,
\end{equation}
or equivalently to the unconstrained linear program
\begin{equation} \label{eq:unconl1Min}
\underset{\bm c}{\textrm{min}} \left \| \bm u-\bm \Psi \bm c \right \|_{2}+ \lambda\left \| \bm c \right \|_{1}, 
\end{equation}
where $\lambda$ is the regularization parameter. The two constrained and unconstrained formulations can be made equivalent, i.e. for every value of $\epsilon$, there is a value for $\lambda$  that makes the two programs equivalent. $\ell_{1}$ minimizations of Eq.~\ref{eq:consl1min} and \ref{eq:unconl1Min} are mostly known as the Basis Pursuit Denoising (BPDN) approach \cite{chen2001atomic}. 

Theoretical results have shown that under certain conditions the solution to the $\ell_{1}$ minimization is indeed the sparsest solution, i.e. it matches the solution to the $\ell_{0}$ minimization \cite{donoho2003optimally,gribonval2003sparse}. However, it is a non-trivial task to guarantee that these conditions are met for a specific problem. As a result, it is generally accepted that the $\ell_{1}$ minimization gives suboptimal solutions. However, due to its computationally appealing convex property, it is still most widely used. Efforts have been made to reformulate the problem such that sparser solutions can be obtained. In particular, non-convex penalty functions that further promote sparsity are shown to outperform $\ell_{1}$-norm. For example, numerical studies have shown that $\ell_{0.5}$ minimization and  $\ell_{1}-\ell_{2}$ minimization problems can lead to a sparser solution than that obtained from an $\ell_{1}$ minimization problem \cite{xu2012regularization, yin2015minimization}.      
 
Regardless of the penalty function that is used to promote the sparsity, the first step in the minimization problem is to calculate the measurement matrix $\bm \Psi$. This requires us to have established the order $k$ for PCE, and have obtained the random input samples $\{\bm \xi^{(i)}\}_{i=1}^{M}$. In this work, we postulate that in PCE models representing physical systems, among all the possible combinations of dimension and order that approximate $u(\bm \Xi)$  within the inaccuracy tolerance $\epsilon$, there exists a PCE that is the sparsest. For instance, this sparsest solution may be corresponding to a sub-dimensional PCE at a high order. Let us denote the order and dimension of this sparsest PCE by $k^{*}$ and $d^{*}\leqslant d $, respectively. It is critical what to set as the dimension and order of the expansion before the sparse recovery. If our choice of PCE truncation order $k$ is in fact smaller than $k^{*}$, we will fail to include sufficient bases in the measurement matrix. As a result, our measurement matrix will not include the bases that the sparsest PCE lives in and thus the compressive sampling solution will not produce the sparsest possible PCE. On the other hand, if the total PCE order $k$ is set  larger than $k^{*}$, we will have a fatter measurement matrix, with extra bases in addition to those that sufficiently span the sparsest possible PCE. A fat measurement matrix  can lead to a large coherence impairing the sparsity of the solution, as will be described in detail later. 

We, so far, have briefly described the implication of PCE order on the recovery of the sparsest solution. The choice of dimension will also have implications. Conventionally, the dimension of PCE, $d$, is the set to be the number of all random inputs to the system. However, in many physical systems, there are  input parameters that have a insignificant influence on the output. If these uninfluential parameters are included, it increases the coherence of measurement matrix without significantly improving the chance of accurate recovery, and thus decreases the solution sparsity. Therefore, it is particularly important in compressive sampling, to include in the PCE as few input parameters as needed given the inaccuracy tolerance $\epsilon$.  In the next sections, we include the theoretical foundations for these arguments and introduce an incremental algorithm that identifies the influential input parameters by searching for the dimensionality of the sparsest solution, $d^{*}$ and also identifies the optimal expansion order $k^*$. 

\section{Efficient recovery of sparse solution}~\label{sec:seq_cs}
In an overdetermined case of estimation chaos coefficients, by increasing the order or dimension of the approximation, the accuracy of the approximate PCE is expected to increase. However, in underdetermined cases, when a sparse recovery is pursued given a fixed number of samples, the accuracy does not necessarily increase when a higher order or higher dimensional expansion is used. That is because the mutual coherence of measurement matrix will increase for expansions with higher order or dimension.  
Coherence or mutual coherence of any matrix $\bm \Psi \in \mathbb{R}^{M \times K}$ is defined as the maximum absolute normalized inner products, i.e., cross-correlation between its columns \cite{doostan2011non, donoho2006stable}.  Let $\psi_{1}, \psi_{2}, ..., \psi_{K} \in \mathbb{R}^{M}$ be the columns of matrix $\bm \Psi$, the coherence of matrix $\bm \Psi$, $\mu(\bm \Psi)$, is then given by
\begin{equation} \label{eq:coherence}
\mu(\bm \Psi):= \underset{1\leqslant  i, j\leqslant  K, i\neq j}{\max}\frac{|\psi_{j}^{T}\psi_{i}|}{\| \psi_{j} \|_{2} \| \psi_{i} \|_{2}}.
\end{equation}

This section follows with mathematical supports on why a large coherence deteriorates the accuracy of compressive sampling solution. It is worth noting that these results apply to measurement matrices that consist of any bases, except for the last result that is only valid for a measurement matrix consisting of PC bases, hereinafter referred to as PC measurement matrix. For notation brevity, however, we use the notation $\bm \Psi$ interchangeably for both general and PC measurement matrices, while making the distinction in the wording of statements.

\subsection{Impact of coherence in sparse recovery}
A large body of work in compressive sampling has focused on the accuracy and stability of this approach. In what follows, we provide a number of theorems developed in this field that are related to coherence.

A widely used property  in theoretical results in compressive sampling is the restricted isometry property (RIP). Specifically, $S$-restricted isometry constant of a matrix $\bm \Psi$ is the smallest $\delta_{S} \in (0,1)$ such that
\begin{equation} \label{eq:RIP}
(1-\delta_{S})\left \| \bm c \right \|_{2}^{2}\leqslant  \left \| \bm \Psi^{S^{*}}\bm c\right \|_{2}^{2}\leqslant  (1+\delta_{S})\left \| \bm  c \right \|_{2}^{2},
\end{equation}
for every submatrix of $\bm \Psi^{S^{*}} \in \mathbb{R}^{M \times S^{*}},  S^{*} \leq S$, of $\bm \Psi\in \mathbb{R}^{M \times K}$ and every vector $\bm c \in \mathbb{R}^{S^{*}}$. 

\begin{theorem}[\cite{cai2010new}] \label{Theory2}
Suppose $\bm u= \bm \Psi \bm c^{*}+ \bm z$ with $\left \| \bm z \right \|_{2}\leqslant  \epsilon$, where $\bm c^{*}$ is S-sparse ($S>1$). If $\delta_{S} < 0.307$, the solution of the $\ell_{1}$ minimization problem given in (\ref{eq:consl1min})
denoted by  $\overline{\bm c}$ satisfies
\begin{equation}
\left \| \bm c^{*}-  \overline{\bm c} \right \|_{2} \leqslant  \frac{\epsilon}{0.307 - \delta_{S}},
\end{equation}
or in noiseless situation exactly recovers $\bm c^{*}$.
 \end{theorem}

 Although Theorem \ref{Theory2} does not directly involves coherence, it is closely related to coherence. The following simple proposition explains this connection. 
\newtheorem{prop}{Proposition}
\begin{prop}[\cite{maleki2010approximate}] \label{prop1}
Suppose that the columns of $\bm \Psi$ are normalized. Then for any $S< \mu(\bm \Psi)^{-1}+1$, $\bm \Psi$ satisfies $\delta_{S}	\leqslant   (S-1)\mu(\bm \Psi)$. 
\end{prop}
Proposition \ref{prop1} makes it obvious that a smaller coherence enforces smaller $\delta_{S}$, thus smaller recovery error. In general, although the RIP-based theorem provides stronger results, it does not offer much applicability. This is because for a given measurement matrix, as opposed to the simplicity in calculating  coherence, calculation of  RIP constant   is known to be an NP-complete problem \cite{maleki2010approximate}.

The first attempt to apply compressive sampling in the estimation of PC coefficients was reported in \cite{doostan2011non}, where conditions on the stability of estimated  coefficients were investigated. Specifically, the following proposition established conditions on the sparsity of PCE solution to ensure stability with a high probability.

\begin{prop}[\cite{doostan2011non}] \label{prop2}
Let $u_{k}(\bm \Xi)$ be a S-sparse Legendre PC approximation of function $u(\bm \Xi)$ with an index set that is a subset of $\Lambda_{k,d}$. Also, let the $\bm c^{*}$ be the vector of PCE coefficients which satisfies the sparsity condition
\begin{equation}
\left \| \bm c^{*} \right \|_{0} = S < 0.25\left(1+\frac{1}{\mu(\bm \Psi)}\right). 
\end{equation}
 Then, with probability larger than $1-\exp\left(-\frac{M}{8(3^{2k})} \right)$ , the $\ell_{1}$ minimization solution, $\overline{\bm c}$, given in (\ref{eq:consl1min}) obeys
 \begin{equation} 
 \mathbb{E}\left [ \left ( u_{k}(\bm \Xi)-\sum_{\bm \alpha \in \Lambda^{k,d}}\overline{ c}_{\bm \alpha} \psi_{\bm \alpha}(\bm \Xi)\right )^{2} \right ] \leqslant  \frac{4}{3M}\frac{(\epsilon + \left \| \bm \Psi \bm c^{*}-\bm u \right \|_{2})^{2}}{1-\mu(\bm \Psi)(4\left \| \bm c^{*} \right \|_{0}-1)}.
 \end{equation}
\end{prop}
 The impact of coherence is transparent in Proposition \ref{prop2}. The smaller the coherence, the more accurately  the  $\ell_{1}$ minimization solution can approximate  the actual sparse solution. 
 
 %$1-\exp(-(-\frac{M}{8K^{\frac{2.2k}{\ln K}}}))$
 
 \subsection{Coherence of a measurement matrix consisting of PC bases}
\begin{figure}[H]
 	\begin{subfigure}[t]{0.32 \linewidth}
 		\includegraphics[width=.94\linewidth]{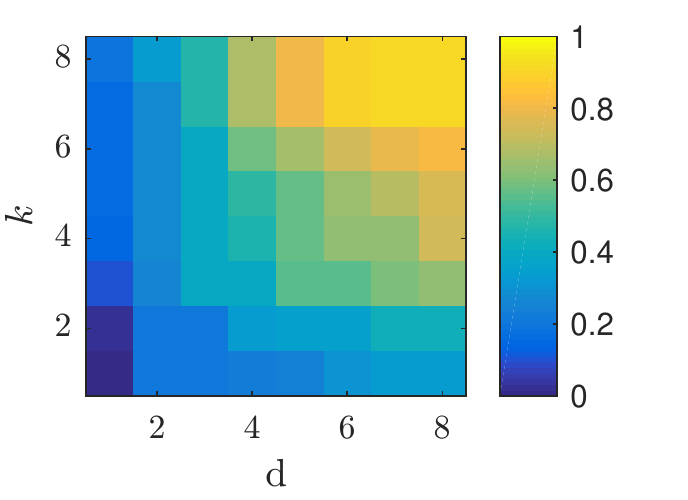}
 		\caption{}	
 		\label{fig:Cha}
 	\end{subfigure}
 	\quad	
 	\begin{subfigure}[t]{0.32 \linewidth}
 		
 		\includegraphics[width=.94\linewidth]{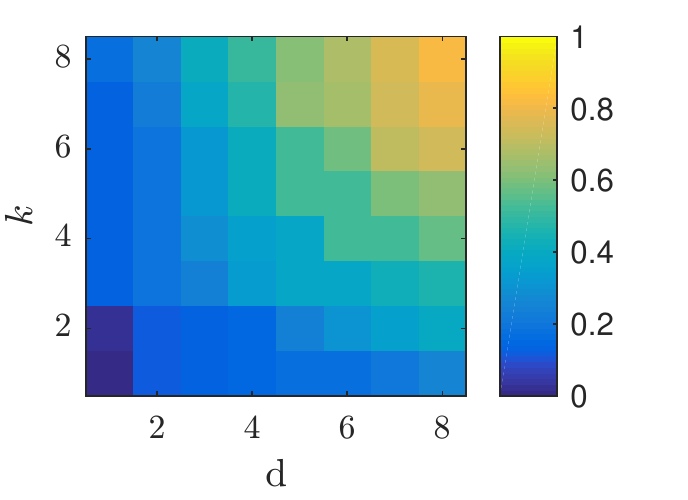}
 		\caption{}
 		\label{fig:Chb}
 	\end{subfigure}
 	\quad
 	\begin{subfigure}[t]{0.32 \linewidth}
 		\includegraphics[width=.94\linewidth]{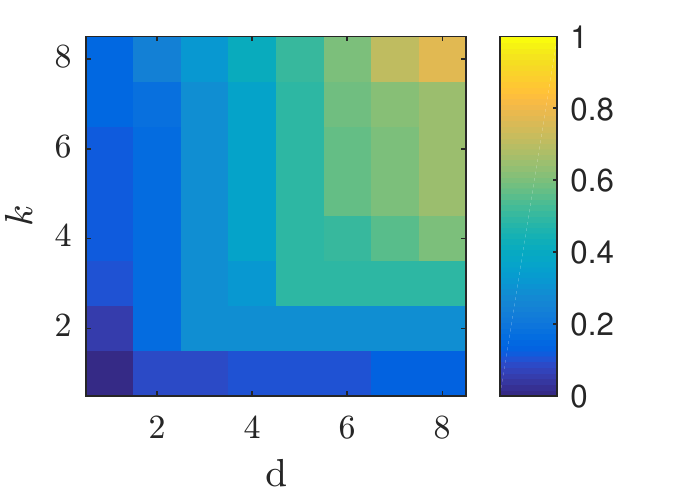}
 		\caption{}	
 		\label{fig:Chc}
 	\end{subfigure}
 	
 	\captionsetup{}
 	\caption{Coherence of PC measurement matrices populated based on evaluations of Legendre polynomials with different combinations of dimension. $d$ and total order, $k$. The polynomials are evaluated at (a) 100 uniform samples, (b) 200 uniform samples, and (c) 300 uniform samples.} 
 	\label{fig.coherence}
 \end{figure}

The PC measurement matrix, i.e. the measurement matrix built by PC bases, is a particular case of measurement matrices, and thus offers particular property. Specifically, let us consider how changes in the construction of the model impact the coherence of measurement matrices. For instance, if a Fourier series expansion is to be estimated, the model design parameter is the set of frequencies. In this case, if a new richer expansion is to be constructed by adding a single new frequency, the number of columns in measurement matrix will be increased by two.  In the case of a PCE model, however, the model design parameters are dimension and order, and a one step increment in each increases the size, and therefore coherence, of the PC measurement matrix, due to Eq.~\ref{eq:NumOfbasis}.  Figure \ref{fig.coherence} shows numerical results highlighting such changes.  Specifically, we have evaluated Legendre polynomials at different combinations of total order and dimension (each varying from 1 to 8), at uniform sample sets of size 100, 200 and 300. It can be seen that the increase in coherence with respect to total order and dimensionality has a higher rate for smaller sample sets.

In models for physical systems, it is often the case that not all the input parameters  have a significant impact on the output. Therefore, when the number of samples is limited and compressive sampling is to be used, including bases corresponding to non-significant random inputs can undesirably increase the coherence of PC measurement matrix, while failing to provide valuable dictionary for better recovery. This can significantly deteriorate the accuracy of coefficient estimation. It should be noted that Theorem\ref{Theory2} holds only when the signal of interest is in fact sparse in the dictionary $\bm  \Psi$.  Therefore, a sparse recovery is successful if the particular choice of dictionary can sparsely represent the signal of interest \cite{aharon2006img}.

 Recently, efforts have been made in learning the dictionaries, i.e. the best bases, in which signals are sparse, as opposed to using predefined and off-the-shelf dictionaries such as wavelet and cosine dictionaries.  Numerical results have shown significant advantages realized by learning the dictionary  \cite{aharon2006img, mairal2009online, mairal2009supervised}.  There are two classes of approaches in dictionary learning: (a)   approaches that identifies the best dictionary among a set of candidate dictionaries \cite{cevher2011greedy}, and (b) more complex approaches that adopt a dictionary from scratch using iterative algorithms \cite{aharon2006img,mairal2009online, mairal2009supervised}.  In compressive sampling for PCE, since the type of bases is predetermined based on the distribution type of inputs, the first class of approaches is applicable. 
In what follows, we  introduce our incremental algorithm which is able to select the right dictionary given a coherence budget dictated by the number of samples. 

 \subsection{A coherence-aware incremental algorithm for sparse PC approximation}
Our objective is to  learn, or intelligently choose, the bases in the measurement matrix or dictionary $\bm \Psi$, given a set of samples. For practical reasons, in working with PCE, we translate this objective into that of identifying the ``ideal" combination of dimensionality and  expansion order. As suggested in \cite{doostan2011non} and is apparent from Figure~\ref{fig.coherence}, in high dimensional PCEs, the expansion order should be kept very small to avoid large coherence for PC measurement matrix. In such cases, if there really exists a sparser PCE approximation at a higher order, it will not be recovered. Our approach to overcome this builds up on (a) the aforementioned argument that in physical systems, input parameters could rank differently based on their impact on the output, and (b) the possibility that by removing uninfluential parameters from the expansion, a higher order sparser PCE, or in general a sparser PCE, is likely to be recovered. In other words, given the competing influence of dimension and order on coherence, we compromise on the former to allow for the latter to increase without largely affecting the coherence. The success of such sparse recovery hinges on how accurate a lower dimensional expansion can still approximate the system of interest. 

To implement this, an incremental selection approach for learning the PC measurement matrix is proposed that identifies the influential input parameters, given the available samples set, and prevents the PCE construction to be unnecessarily high dimensional. We  postulate that given a  tolerance budget, there exists a sparsest PCE approximation among various combinations of dimension and order levels. To reach this combination, we develop an incremental search whose steps involve one unit trial increases in a dimension or the order. For each trial increment, in dimension or order, a set of polynomial bases at once enter the measurement matrix. To determine which trial should be selected at the end of each step, the  following two assumptions furnish the foundation for our approach. 

\begin{assumption}\label{assump1}
 If the trial basis set added to the measurement shares bases with those spanning the sparsest PCE approximation, the sparsity of the resulting $\ell_{1}$ minimization solution will increase. 
\end{assumption}
\begin{assumption}\label{assump2}
Once all the bases that span the sparsest approximation are included in the measurement matrix, adding new bases will only increase the coherence of  measurement matrix and can therefore impair the sparse recovery.  
\end{assumption}

Assumption~\ref{assump2} is always true, while Assumption~\ref{assump1} may not be true if the additional trial bases increase the coherence of measurement matrix so drastically, that even though informative  bases are provided, a sparser solution can still not be recovered. However, this is typically not the case in our incremental algorithm, where trial increments in coherence in each step is usually small.  
 
Based on these two assumptions, the incremental algorithm searching through different dimensions and order levels is developed as follows. At the beginning of step $t$ in this search, let $\bm \Xi^{r}_t$ denote a reduced dimension random vector which already  includes $d_t$ significant random inputs, i.e. $\vert \bm \Xi^{r}_t \vert = d_t$, $d_t < d$ and $\vert \cdot \vert$ loosely denotes the dimension of a random vector. Given this reduced dimension, and an expansion order $k_t$, the PCE expansion at the beginning of step $t$ is given by
\begin{equation}
u_{k_t,d_t}(\bm\Xi^r_t) := \sum_{\bm \alpha \in \Lambda_{d_t,k_t}} c_{\bm \alpha } \psi_{\bm \alpha}(\bm \Xi^r_t).
\end{equation}
Driven by the two assumptions, the algorithm keeps incrementing the dimension or the order of PCE depending on how the sparsity of $\ell_{1}$ minimization solution is improved, as will be explained in detail. The incremental search algorithm stops when  sparsity can no longer be improved. At each step, the algorithm chooses among (a) adding one of the parameters to $\bm \Xi^{r}_{t}$, (b) increasing $k_{t}$ by one, or (c) termination. As mentioned earlier, the single selection criteria is the sparsity of the solution for each trial increment compared to the sparsity at the previous step. 

Specifically, at step $t$ of the search, each one of the remaining random inputs, i.e. those that are not already included in $\bm \Xi^{r}_{t}$, is added to $\bm \Xi^{r}_{t}$ once at a time and $\ell_{1}$ minimization is solved and the sparsity of the solution is recorded. An $\ell_{1}$ minimization is also solved for the $\bm \Xi^{r}_{t}$ of the previous step with an incremented order $k_{t}$. If the sparsity of the solution is not improved in either of these cases, the algorithm stops. Otherwise, if adding the incremented order yields the sparsest solution, then the new sparsest PCE solution at the next step will be $u_{k_t+1,d_t}(\bm\Xi^r_{t+1})$. If the sparsest solution is caused by the addition of a new input parameter, then that random input is added to $\bm \Xi^{r}_{t}$, and the new sparsest PCE solution will be recorded as  $u_{k_t,d_t+1}(\bm\Xi^r_{t+1})$. 

Algorithm \ref{pseudocode_sparsity} shows the pseudocode for  our incremental algorithm. The algorithm is initialized with a zero-dimensional 2nd order expansion. Instead of order 2, larger values can also be used depending on the prior knowledge about the true order of the solution. Similarly, if a number of  influential  inputs are \emph{a priori} known, they can be included in the initial $\bm \Xi^{r}_{t=1}$.

\begin{algorithm}[H]
	\caption{Incremental algorithm for PCE coefficients estimation}\label{pseudocode_sparsity}
	\begin{algorithmic}[1]
		\State Set $t=1$, $\vert\bm \Xi^{r}_{t}\vert = 0$, $k_{t}=2$. \Comment{Start with a zero-dimensional 2nd order expansion.}
		\State Set $s^*=0$ and $s_{min}$ to be a large number \Comment{$s$ denotes the number of non-zero coefficients.}
		\While {sparsity is improving  ($s^* < s_{min}$)}
		\For {$\{ \Xi_i: \Xi_i \not\in \bm \Xi^{r}_{t}\}  $}
		\State Temporarily add input parameter $\Xi_{i}$, to $\bm \Xi^{r}_{t}$. \Comment{$\vert\bm \Xi^{r}_{t}\vert = \vert\bm \Xi^{r}_{t}\vert + 1$}
		\State Set $\bm c^*_i$ to be the solution of related Eq.~\ref{eq:consl1min}.
	    \State Remove input parameter $\Xi_{i}$ from $\bm \Xi^{r}_{t}$.
		\EndFor
		\State Set $s^{(d)}$ to be  $\ell_0$-norm of the sparsest recorded solutions $\bm c^*_i$. 
	    \State Temporarily add $k_{t}$ by one, solve the related Eq.~\ref{eq:consl1min}, and record its solution's $\ell_0$-norm as $s^{(k)}$.
		\State $s^* = \min \{s^{(k)} , s^{(d)}\}$   \label{k_or_d} \Comment{Pick the  sparser solution.}
		\If {$s^* < s_{min}$} \Comment{Check if the new solution is sparser than the previous step.}
		\State Increment the dimension or order based on Line~\ref{k_or_d}.
		\State $s_{min} = s^*$.
		\State $t=t+1$.
		\EndIf
		\EndWhile
		
	\end{algorithmic}
\end{algorithm}

At initial steps of the sequence, i.e. for a small $t$ and when $d_{t}$ and $k_{t}$ are also small,  the number of available samples, $M$, can be larger than the number of unknown chaos coefficients, $\vert \Lambda_{d_t,k_t} \vert$. In such overdetermined cases, to decide whether the increment should be in certain dimension or in the order, instead of inspecting changes in sparsity, we inspect changes in  least square error (LSE). It has been shown that  that minimizing the least square error  is an effective approach for basis selection \cite{blatman2010adaptive}. To decide where to switch the criterion from LSE to sparsity, we make use of the desired accuracy value, $\epsilon$ in Eq.~\ref{eq:consl1min}, and while LSE is larger than this desired value, instead of solving $\ell_{1}$ minimization and recording the sparsity, we solve the following minimization
\begin{equation} \label{eq:minMSE}
\underset{\bm c}{\textrm{min}}\left \| \bm u -\bm \Psi \bm c \right \|_{2}^{2}.
\end{equation}

\subsection{Note on computational cost}
The proposed incremental approach involves solving Eq.~\ref{eq:minMSE} and Eq.~\ref{eq:consl1min} multiple times for PCEs with different dimension and order. This, however, does not result in a substantial computational cost as these equations can be solved with very efficient algorithms. At initial steps, where  $\ell_2$ minimization problem in Eq.~\ref{eq:minMSE} is solved, the computational cost is of $\mathcal{O}(MK^{2})$, mainly due to matrix-matrix multiplication. It should be noted that  initial steps are applied to low dimensional low order expansions where $K$ is  small, and in cases which does not involve large number of samples, the algorithm soon switches to $\ell_1$  minimization of Eq.~\ref{eq:consl1min}. To solve this $\ell_1$ minimization problem, BPDN  solver in \verb SPGL1  package \cite{van2007spgl1}, can be used, which runs a fast algorithm with computational cost that is mainly due to matrix-vector multiplication with complexity of $\mathcal{O}(M K)$ \cite{van2008probing}. 

The final cost of incremental algorithm depends on the number of steps before reaching the sparsest solution,  which depends on $d^*$ and $k^*$. Therefore, the additional  cost incurred by this incremental approach will be moderate when only a small subset of random parameters are in fact influential. As an example, in an 80-dimensional problem, which is described in detail in Section~\ref{sec:highdimExample}, where $d^*=5, k^*=3$, and $M=500$, the incremental algorithm took only 50.56 sec on a personal computer. It should be noted that in achieving the final goal of constructing a PCE to approximate a complex model, this additional cost  is typically negligible compared to the large sampling cost imposed by expensive simulations of the model under study. Also, the additional cost should be considered in conjunction with the  significant benefits materialized as a result of a more accurate approximation and a lower dimensional representation.

In next section, we provide the results of numerical validation for the proposed algorithm. Hereafter, we refer our algorithm as \emph{incremental}, and refer to $\ell_{1}$ minimization with a full expansion as \emph{conventional}. 

\section{Numerical examples}~\label{sec:numericalresults}
In this section we provide four different numerical examples to demonstrate the validity of our proposed algorithm and compare it with other approaches.  Uniformly distributed random variables together with Legendre polynomials are used in all examples. In the first example, the  target function is a manufactured sparse low dimensional polynomial expansion with arbitrary coefficients. In the second example, the solution of a stochastic diffusion problem is considered. The third example concerns approximation of a sparse high dimensional polynomial expansion. Finally, in the fourth example, we address approximating community land model using our proposed algorithm. The first three examples are designed in a way that a subset of input parameters are influential. The forth example, on the other hand, is a high dimensional real world problem where the number of influential random input is not known.  We have used the MATLAB package \verb SPGL1 \cite{van2007spgl1} to  solve the $\ell_{1}$ minimization problem in~\ref{eq:consl1min}  in the conventional approach and also at each step in our incremental algorithm. We have used the MATLAB function \verb lsqlin  for LSE minimization in the initial steps of the incremental algorithm. In all examples, we start the incremental algorithm with $k_{1}=2$ and set $\bm \Xi^{r}_{1}$ to be an empty vector.

\subsection{Example 1: A low dimensional manufactured polynomial expansion} \label{sec:manufacPCE10d}
\begin{figure}	
 	\centering
 	\begin{subfigure}[t]{\linewidth}
 		\centering
 	 		\includegraphics[width=.75\linewidth]{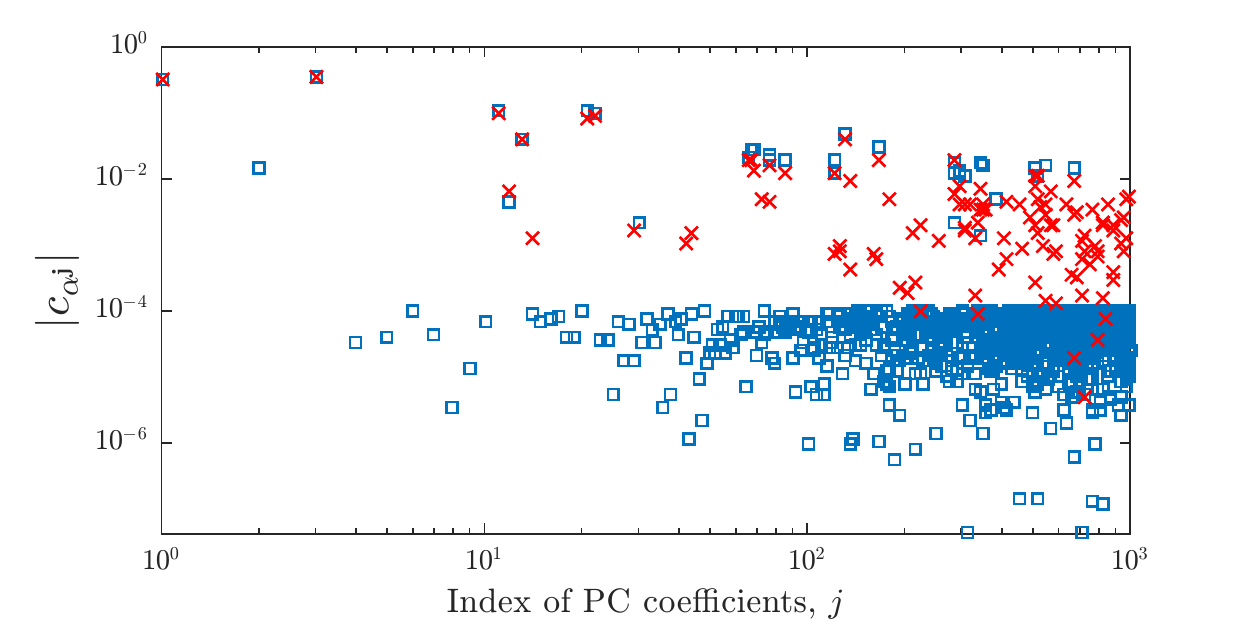}
 	 		\caption{Conventional approach with $d=10$ and $k=4$} \label{fig:coeffc}
 	\end{subfigure}
 	\\
 	\begin{subfigure}[t]{\linewidth}
 		\centering
 		\includegraphics[width=.75\linewidth]{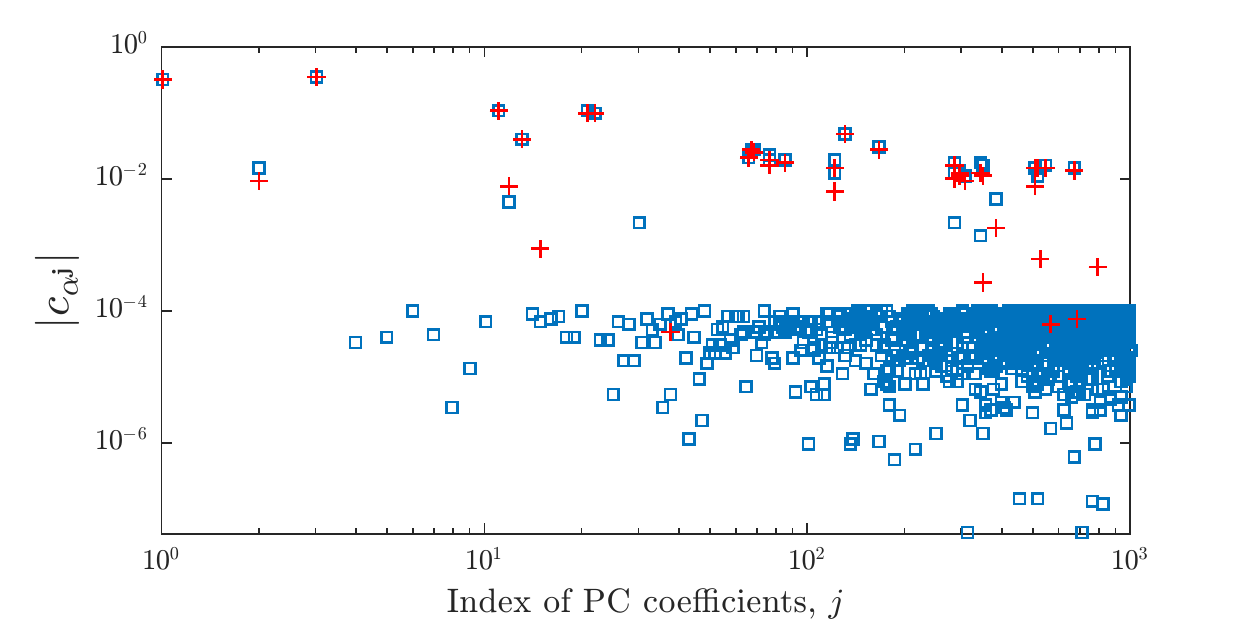}
 		\caption{Conventional approach with $d=5$ and $k=5$} \label{fig:coeffb}
 	\end{subfigure}
 	\\
 	 	\begin{subfigure}[t]{\linewidth}
 	 		\centering
			\includegraphics[width=.75\linewidth]{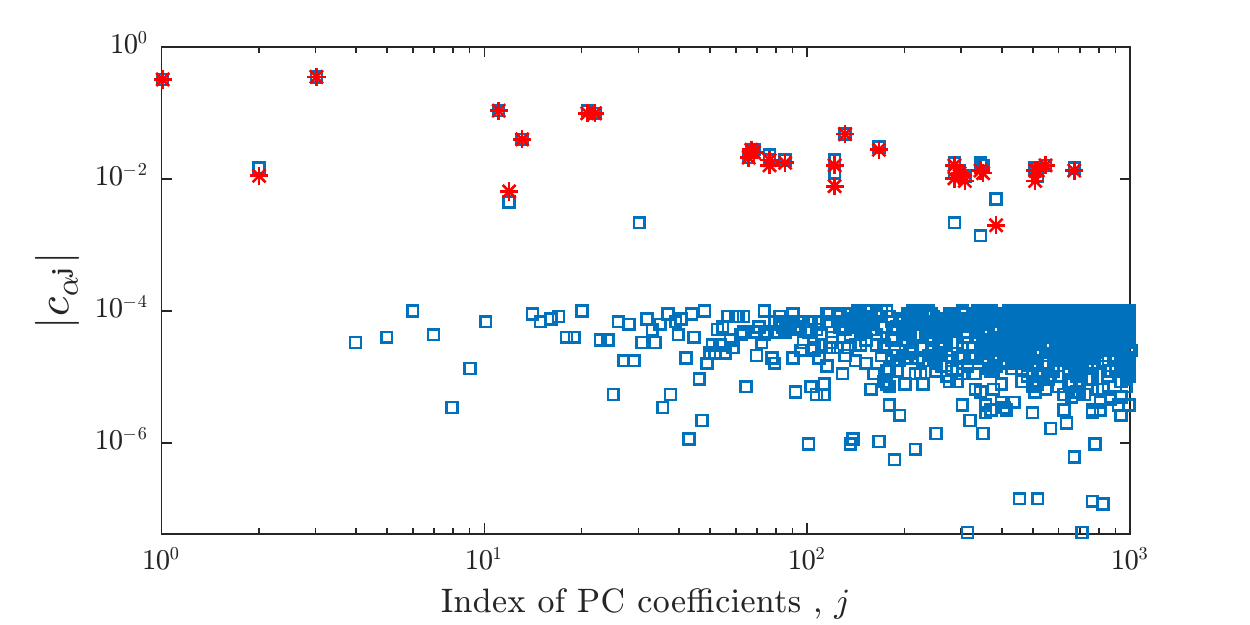}
 		\caption{Incremental approach} \label{fig:coeffa}	
 	 	\end{subfigure}
 	 	\captionsetup{}
 	\caption{Comparison between  true and recovered coefficients for the 4th-order 10-dimensional manufactured polynomial expansion of Section~\ref{sec:manufacPCE10d}. The target function is manufactured so that it can be effectively approximated using a 4th order 3-dimensional expansion. True coefficients are shown by squares.}\label{fig.coefficients.ex1}
 \end{figure}

In this example, we manufacture a target function as a 4th order Legendre polynomial expansion in a 10-dimension random input with uniform density on $[-1, 1]^{10}$. This example is designed to evaluate the performance of different algorithms when the output PCE in fact lives in a smaller dimension than that of the input parameter space. Accordingly, we specify values for the 1001 chaos coefficients such that the target polynomial expansion has significant coefficients in 3 dimensions. Specifically, for each coefficient $c_{\bm \alpha}$, if it only includes any of the first three random inputs, we assign a random realization from $\text{U}(-e^{-\left \| \bm \alpha \right \|_{1}}, e^{-\left \| \bm \alpha \right \|_{1}})$. Otherwise, we assign to all remaining coefficients a random realization from $\text{U}(-10^{-4}, 10^{-4})$. A visualization of the coefficient values is shown as squares in Figure~\ref{fig.coefficients.ex1}.

Our task is to evaluate the performance of our incremental and also the conventional algorithm in recovering the expansion coefficients, if a small number of samples from this manufactured expansion are provided. We set $\epsilon=0.01 \left \| \bm u \right \|_{2}$. For conventional approach, we consider two cases: (a) when there is complete ignorance about the influential inputs and the measurement matrix is evaluated for $d=10$, and (b) when out of the seven insignificant inputs, five are known to the modeler and as a result the measurement matrix is evaluated for $d=5$. The choice of order depends on the number of available samples and also the dimensionality of the expansion. The order must be chosen large enough to ensure the desired accuracy, but nonetheless small enough to prevent the problem from being \textit{heavily} under-sampled and the measurement matrix from being highly coherent. However, there exists no rigorous criterion to determine \emph{a priori} when a problem is \textit{heavily} under-sampled, or in other words, what is the relationship between the number of samples and the largest $k$ that can be used. Prior knowledge on the smallest order that provides the desire accuracy doesn't exist, either. Therefore, the choice of order is typically ad hoc. In this example, for the 10-dimensional case we set $k=4$, while for the 5-dimensional case, which may allow for a higher order, we set $k=5$. 
 
Figure \ref{fig.coefficients.ex1} compares the true coefficients with the recovered ones for a single trial using 100 samples. It can be seen that conventional algorithm results in a less sparse solution. The larger number of non-zero coefficients, in this case, is the direct result of  large coherence of the PC measurement matrix populated in a conventional way with high dimension and order. Namely, when the columns in the measurement matrix are closely related, it becomes less easy to distinguish which basis in fact impact the output \cite{candes2011compressed}; hence a convoluted attribution of impacts.

\begin{figure}

		\begin{subfigure}[t]{0.49 \linewidth}
			\includegraphics[width=.95\linewidth]{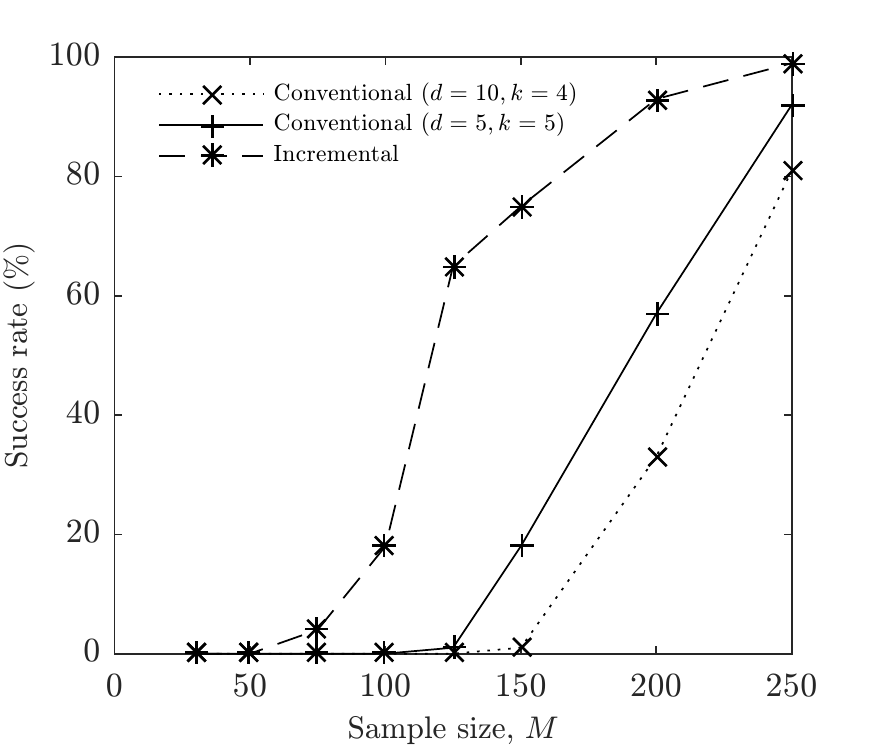}
			\caption{}
			 \label{fig:PCEa}		
		\end{subfigure}
		\quad
		\begin{subfigure}[t]{0.49 \linewidth}
	\includegraphics[width=.95\linewidth]{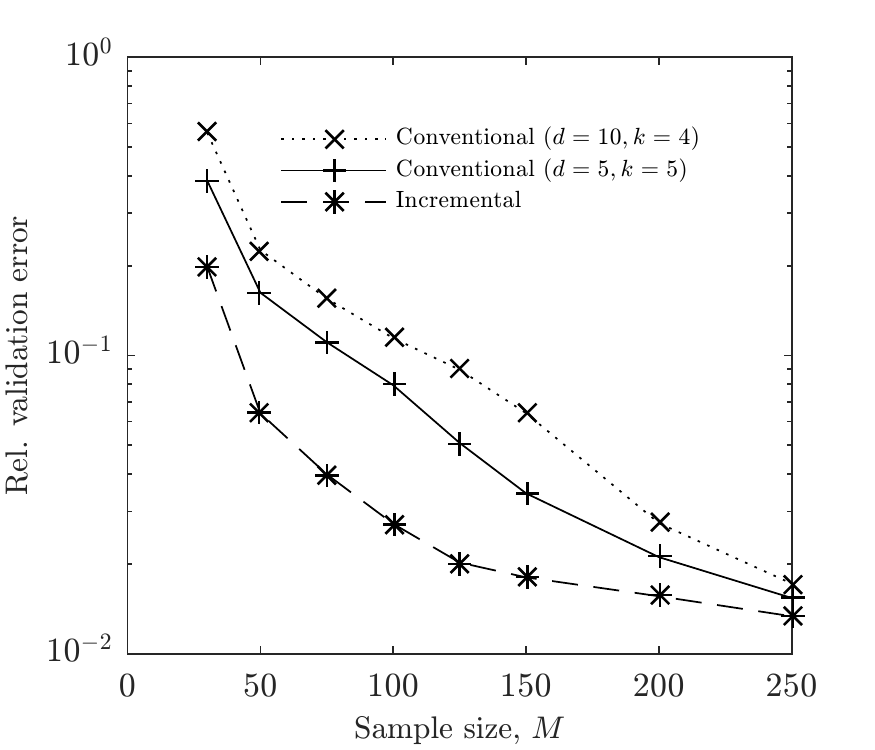}
			\caption{}
			 \label{fig:PCEb}
		\end{subfigure}
		\\
		\begin{subfigure}[t]{0.49 \linewidth}
	\includegraphics[width=.95\linewidth]{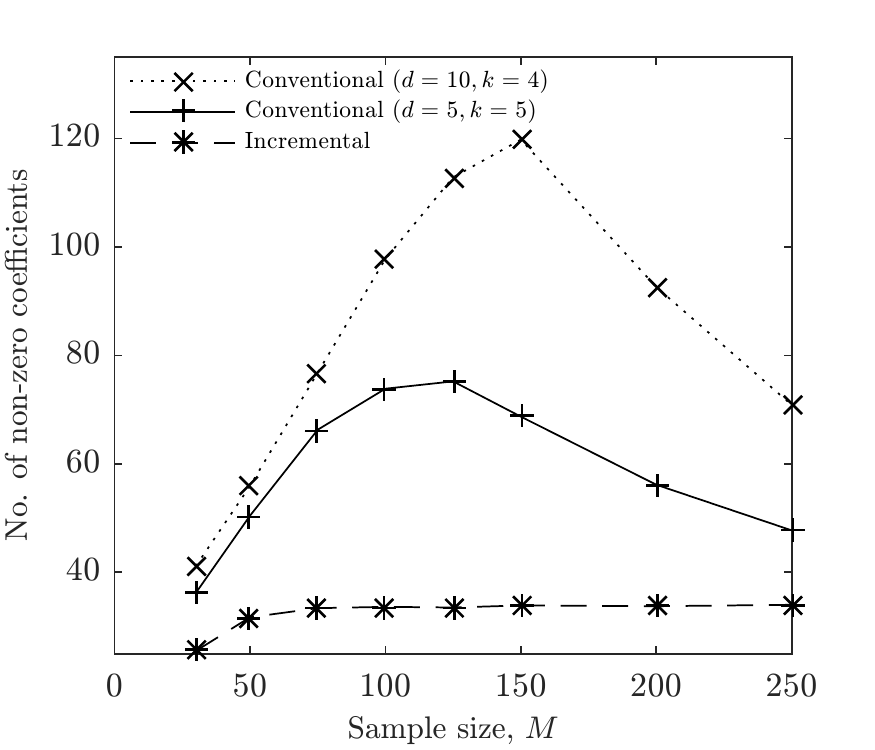}
		\caption{}	
			 \label{fig:PCEc}
		\end{subfigure}
		\quad
		\begin{subfigure}[t]{0.49 \linewidth}
	\includegraphics[width=.95\linewidth]{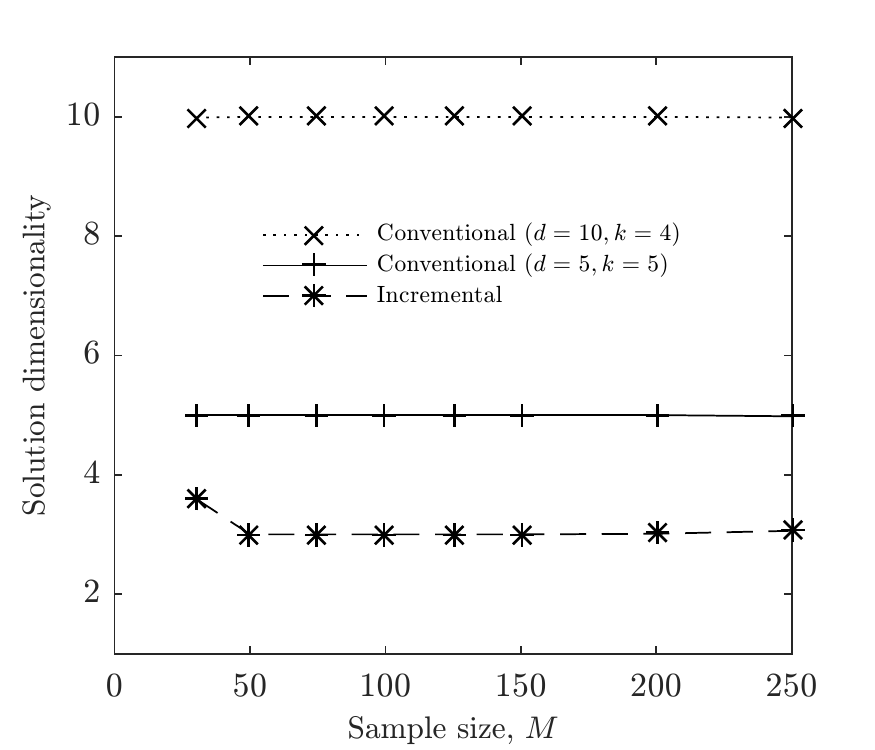}
		\caption{} 
			\label{fig:PCEd}
		\end{subfigure}
\captionsetup{}
	  \caption{Comparison of different performance metrics for incremental and conventional algorithms for the 4th-order 10-dimensional manufactured polynomial expansion of Section~\ref{sec:manufacPCE10d}. Markers show values for (a) success rate (b) relative validation error (c) number of non-zero coefficients (d) solution dimensionality, averaged over 100 independent trials for each sample size.} 
	\label{fig.example1}
\end{figure}

Figure \ref{fig:PCEa} compares the success rate of coefficient recovery for incremental and conventional approaches. A successful recovery in this example is defined to be one that satisfies $\left \|  \overline{\bm c} - \bm c^{*}\right \|_{2} / \left \| \bm c^{*} \right \|_{2} \leqslant  0.02$, where $\overline{\bm c}$ and $\bm c^{*}$ are the vectors of recovered and exact coefficients, respectively. Figure \ref{fig:PCEb} compares relative validation error, defined as $\left \|  \overline{\bm u} - \bm u\right \|_{2} / \left \| \bm u \right \|_{2}$, where $\bm u$ includes evaluations of target (exact) expansion  at 200 new samples from the inputs and $\overline{\bm u}$ includes the  evaluations of recovered expansion at the same 200 sampled inputs. Figure \ref{fig:PCEc} compares the number of non-zero terms in recovered solutions, and finally \ref{fig:PCEd} compares the dimension of recovered polynomials.  In these figures, for a more robust comparison, for each sample size $M$, 100 independent sample sets are generated, based on which the shown values are average quantities over 100 recovery trials. It can be seen that incremental algorithm significantly outperforms  conventional algorithm by uncovering of a lower dimensional structure. Besides, Figure \ref{fig.example1} also suggests that when the predetermined dimension and order in conventional approach are set closer to the ``significant" dimension and order, better results can be obtained.

\subsection{Example 2: A stochastic partial differential equation}
 In this example, we consider the solution of the following stochastic linear elliptic partial differential equation (PDE)
 \begin{equation}\label{PDE}
 	\begin{aligned}
\nabla \cdot (a(\bm x,\bm \Xi) \nabla u (\bm x,\bm \Xi)) &=1, \quad \bm x \in \mathcal{D},
\\
u(\bm x,\bm \Xi) &=0, \quad  \bm x \in \partial \mathcal{D},
 	\end{aligned}
 \end{equation}
 where $\mathcal{D}= (0,1) \times (0,1)$ with boundary $\partial \mathcal{D}$, and $a$ is the random diffusion coefficient given by
  \begin{equation}\label{acoeff}
  	a(\bm x, \bm \Xi) = a_{0} + \sigma \sum_{i=1}^{d} \sqrt{\lambda_{i}}\phi_{i}(\bm x) \Xi_{i},
  \end{equation}
  where $a_{0}=1$, $\sigma = 0.3, d=20$, $\left \{ \Xi_{i} \right \}_{i=1}^{d}$, are uniform random variables each taking values in $[-1,1]$, and $\left \{ \lambda_{i} \right \}_{i=1}^{d}$ and $\left \{ \phi_{i}(\bm x) \right \}_{i=1}^{d}$ are the $d$ largest eigenvalues and  eigenfunctions of the exponential kernel
  \begin{equation}\label{kernelfunction}
  	C(\bm x, \bm y) =  \text{exp}\left [ \frac{-|x_{1}-y_{1}|- |x_{2}-y_{2}|}{l} \right ],
  \end{equation}
  with correlation length $l=1$.
 
\begin{figure}
		\begin{subfigure}[t]{0.49 \linewidth}
			\includegraphics[width=.94\linewidth]{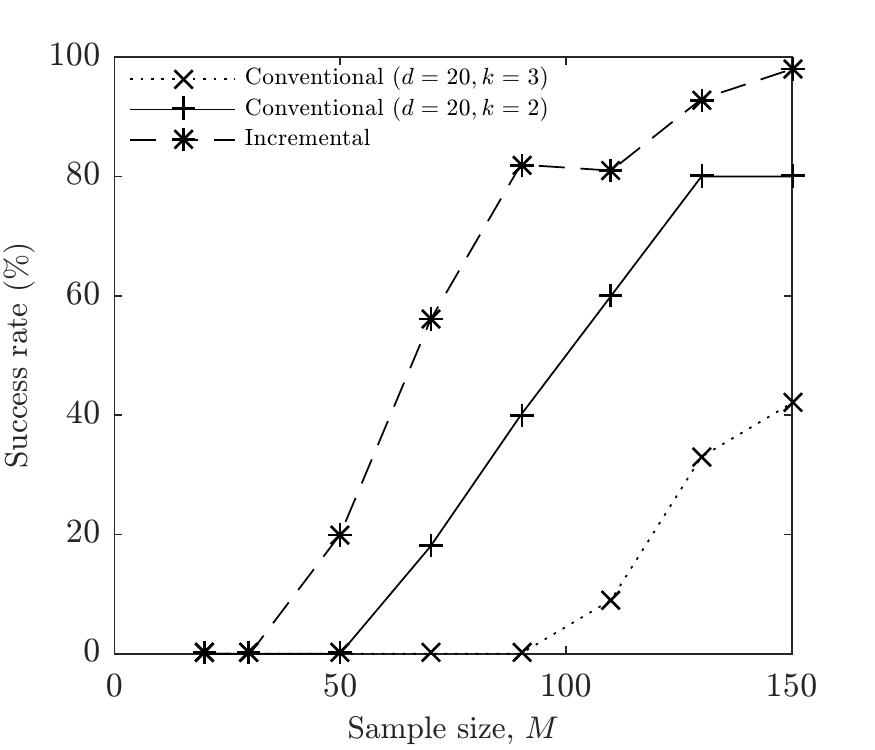}
			\caption{}	
			\label{fig:pdea}
		\end{subfigure}
	\quad	
	\begin{subfigure}[t]{0.49 \linewidth}
		
		\includegraphics[width=.94\linewidth]{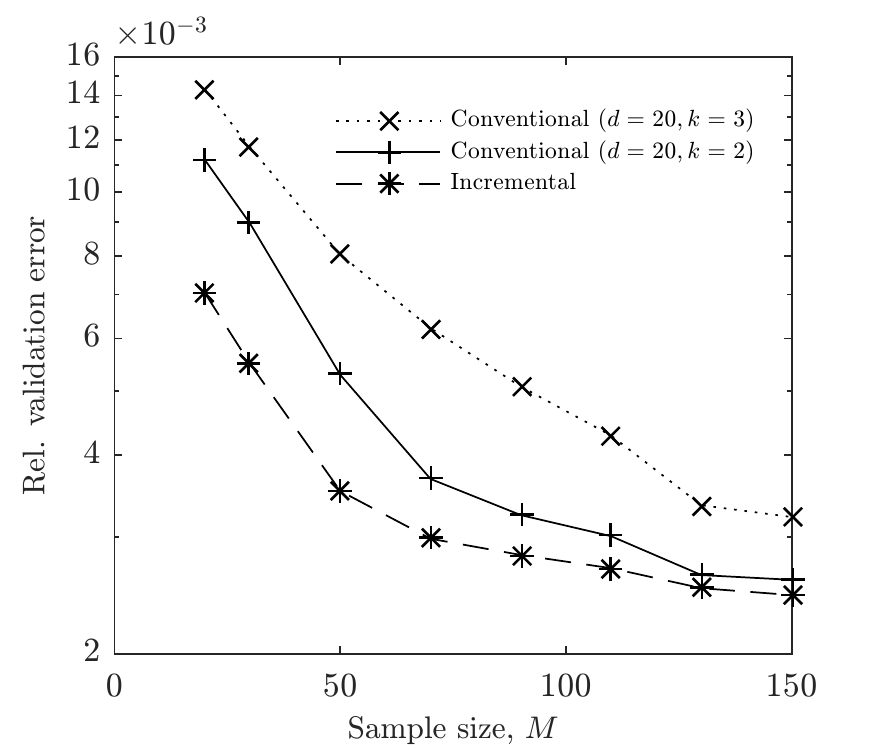}
		\caption{}
		\label{fig:pdeb}
	\end{subfigure}
	\quad
	\begin{subfigure}[t]{0.49 \linewidth}
		\includegraphics[width=.94\linewidth]{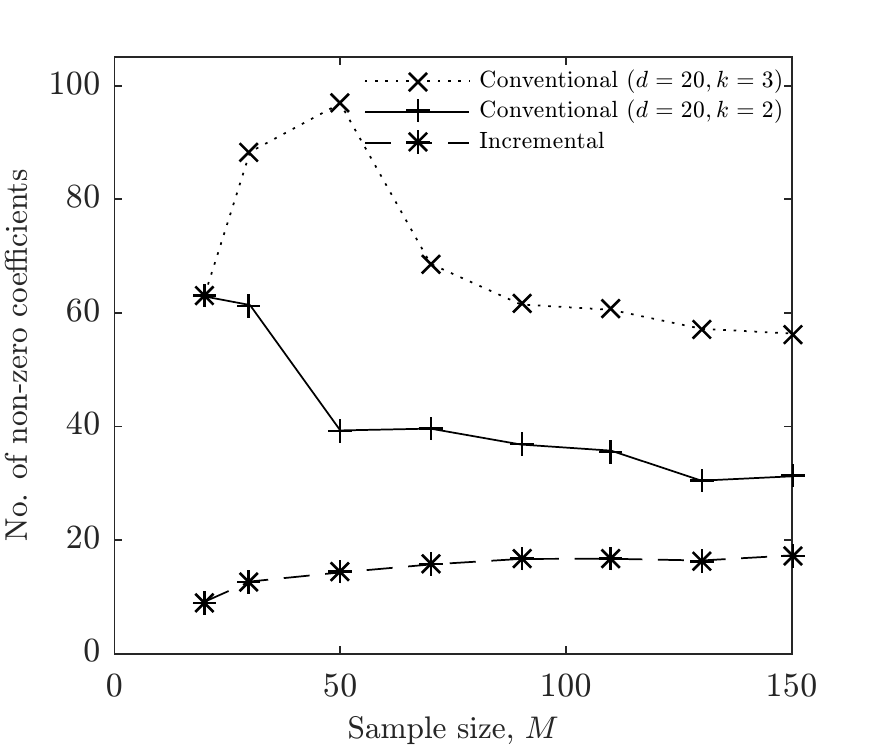}
		\caption{}	
		\label{fig:pdec}
	\end{subfigure}
	\quad
	\begin{subfigure}[t]{0.49 \linewidth}
		\centering
		\includegraphics[width=.94\linewidth]{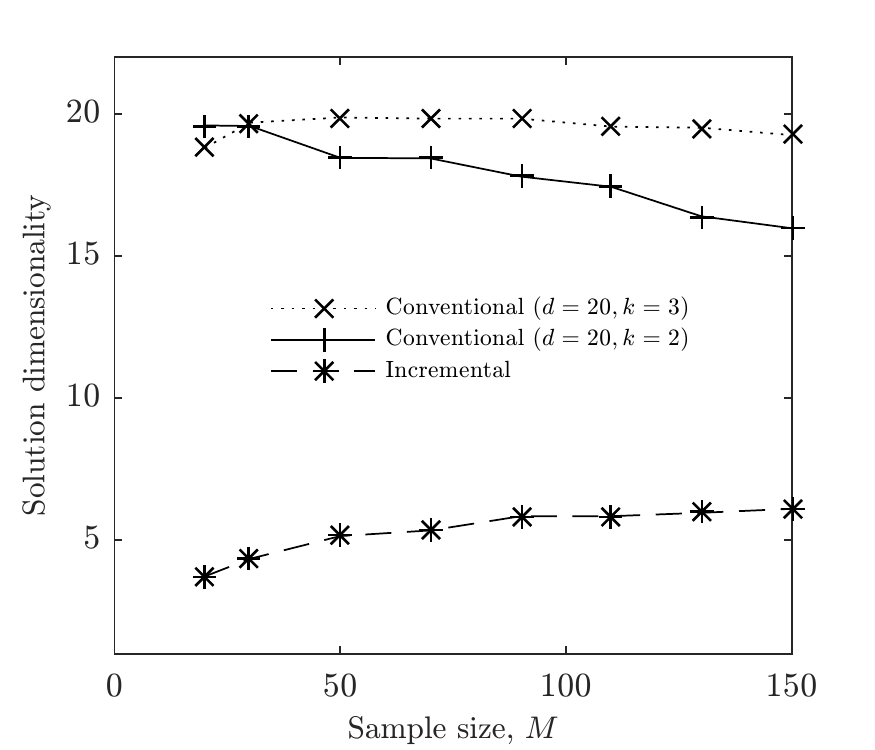}
		\caption{} 
		\label{fig:pded}
	\end{subfigure}
		\captionsetup{}
		\caption{Comparison of incremental and conventional algorithms results for stochastic PDE (a) relative validation error (b) number of non-zero coefficients (c) solution dimensionality.} 
		\label{fig.example2}
	\end{figure}

 For each sample of $\bm \Xi$, we evaluate $u((0.61,0.58),\bm \Xi)$ in Eq.~\ref{PDE} using MATLAB's \verb pdenonlin  as the quantity of interest (QoI). We use our incremental approach together with the conventional compressive sampling approach with $\epsilon= 0.002\left \| \bm u \right \|_{2}$ to obtain PCE approximations of  QoI. In this example, since we do not prior knowledge about $d^*$ and $k^*$, we include all random inputs in evaluation of measurement matrix for conventional approach. Also, since $d$ is large, we need to choose a small expansion order $k$. For a better comparison between conventional and our incremental algorithms, we run the conventional algorithm for two different $k$ values, $k=2$ and $k=3$. Figure \ref{fig.example2} compares the average results for 100 independents trials for incremental and conventional algorithms. Figure \ref{fig:pdea} compares the success rates. We consider $\overline{\bm c}$ to be a successful approximation of PCE coefficients if it satisfies $\left \|  \bm \Psi \overline{\bm c} - \bm u\right \|_{2} / \left \| \bm u \right \|_{2} \leqslant  0.003$, where $\bm u$ is a vector of evaluation of target function at 200 new ``test" samples. Figure \ref{fig:pdeb} compares relative validation errors for the 200  test samples. Figure \ref{fig:pdec} compares the number of non-zero terms in recovered solutions, and finally \ref{fig:pded} compares the dimension of solution PCE after the sparse recovery. As is evident by the results, the incremental algorithm outperforms the conventional algorithm in terms of sparsity and accuracy by exploring the reduced dimension PCE approximation of target function. It can be seen that these improvements in the results are more significant when the sample size is small and as a result coherence is relatively larger for the measurement matrix with prefixed $k$ and $d$. In this example, $k^*$ is found by the incremental approach to be 2. We observe that the closer the prefixed $p$ and $d$ are to the $p^*$ and $d^*$, the better  conventional algorithm can recover the sparsest solution. However, this may not be the case all the time, and therefore the incremental algorithm should be preferred. 
 
\subsection{Example 3: A high dimensional manufactured polynomial expansion} \label{sec:highdimExample}
In the previous two examples, the prefixed expansion order, $k$, in conventional approach was equal or greater than the order of the underlying solution, $k^*$. Therefore, although the measurement matrix in conventional approach suffered from large coherence, it did contain all the bases that the PCE approximation was sparse in. In other words, the conventional approach was ineffective only because it could not identify the few important bases out of a wide set of choices.  In this example, we consider a different class of problems where conventional approach can be ineffective. Specifically, we consider problems whose high dimensionality force the prefixed expansion order $k$ in conventional approach to be set to a small value, likely to be lower than the underlying solution order $k^*$. As a result, the measurement matrix in conventional approach will lack  important higher order bases. Therefore, the conventional approach not only suffers from a large coherence and poor sparse recovery, but also from the inability to form a sufficiently wide set of candidate bases to be identified through the recovery process.

\begin{figure}
	\begin{subfigure}[t]{0.49 \linewidth}
		\includegraphics[width=.94\linewidth]{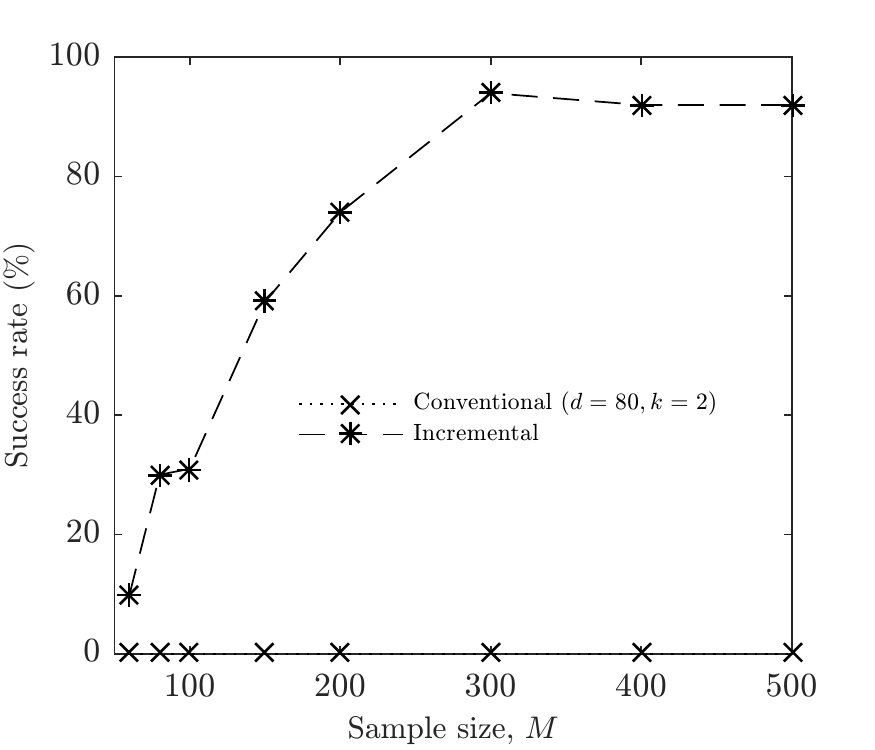}
		\caption{}	
		\label{fig:HDa}
	\end{subfigure}
	\quad	
	\begin{subfigure}[t]{0.49 \linewidth}
		
		\includegraphics[width=.94\linewidth]{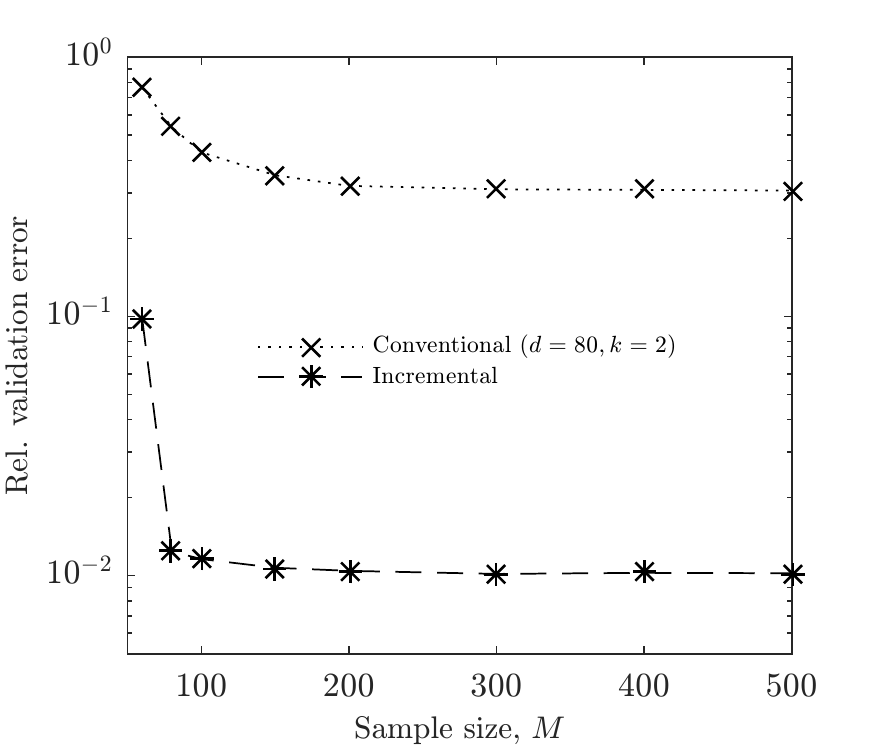}
		\caption{}
		\label{fig:HDb}
	\end{subfigure}
	\quad
	\begin{subfigure}[t]{0.49 \linewidth}
		\includegraphics[width=.94\linewidth]{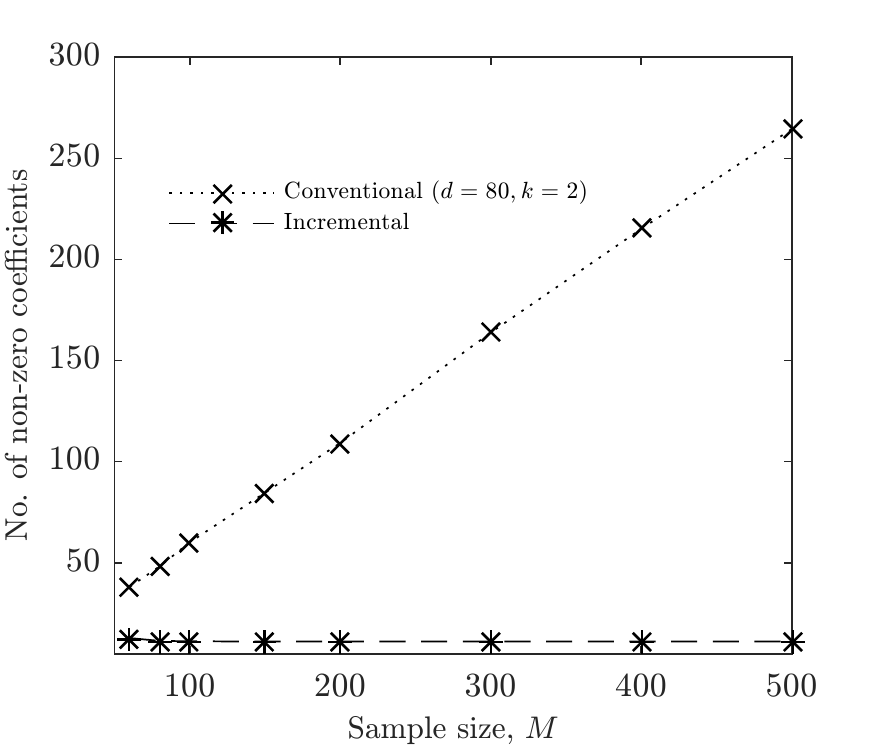}
		\caption{}	
		\label{fig:HDc}
	\end{subfigure}
	\quad
	\begin{subfigure}[t]{0.49 \linewidth}
		\centering
		\includegraphics[width=.94\linewidth]{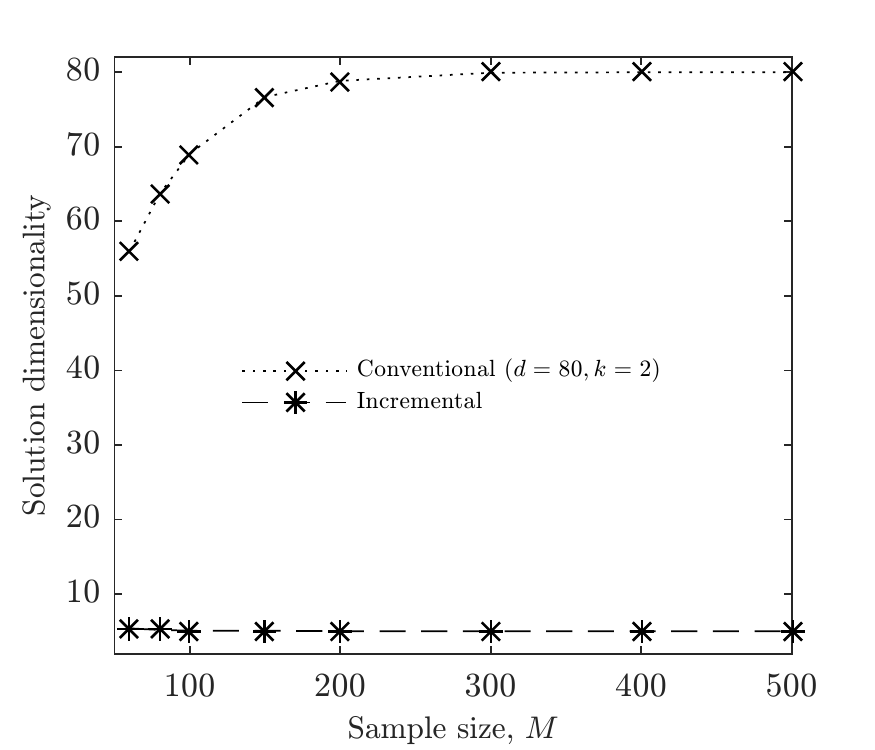}
		\caption{} 
		\label{fig:HDd}
	\end{subfigure}
	\captionsetup{}
	\caption{Comparison of incremental and conventional algorithms results for manufactured high dimensional PCE (a) relative validation error (b) number of non-zero coefficients (c) solution dimensionality.} 
	\label{fig.HD_example}
\end{figure}

To demonstrate the advantage of the incremental approach, let us consider the target function, $u(\bm \Xi)$, to be a third order 80-dimensional polynomial expansion given by
\begin{equation}
	u(\bm \Xi) = \sum_{i=1}^{K_1} \Xi_i + \sum_{i=1}^{K_1-1} \Xi_i \Xi_{i+1} +\sum_{i=1}^{K_{1}-2} \Xi_i \Xi_{i+1} \Xi_{i+2}+ \sum_{i=K_1+1}^K  \frac{1}{5(1+i)^2}\Xi_i+ \sum_{i=K_1+1}^{K-1}\frac{1}{3(1+i)^2}\Xi_i \Xi_{i+1},
\end{equation}
where $K_{1}=5$ and $K=80$ and $\bm \Xi$ is the vector of uniformly distributed random inputs with values in $[-1,1]^{80}$. Evaluations of this target function at sample points in the parameter space are used as observation samples in the sparse recovery algorithms.  For the conventional approach, we set $d=80$ and $k=2$, and note that a third order expansion would lead to a measurement matrix with $91,881$ basis functions and the problems becomes drastically under-sampled. Also, we set $\epsilon= 0.01\left \| \bm u \right \|_{2}$. As expected, the incremental algorithm results in $k^*=3$. Figures \ref{fig.HD_example} compares the average results for 100 independents trials for incremental and conventional algorithms. Figure \ref{fig:HDa} compares the success rates. We consider $\overline{\bm c}$ to be a successful approximation of PCE coefficients if it satisfies $\left \|  \bm \Psi \overline{\bm c} - \bm u\right \|_{2} / \left \| \bm u \right \|_{2} \leqslant  0.011$, where $\bm u$ is a vector of evaluations of target function at 200 new samples. It can be seen that the success rate remains at zero for conventional approach as it never achieves the desired accuracy. Also, in Figure \ref{fig:HDb}, where the relative validation errors for 200 independent test samples are shown, it can be seen that conventional approach results in very large prediction error. This is a direct result of using a measurement matrix that does not include important higher order bases. Furthermore, it can be seen in Figures \ref{fig:HDc} and \ref{fig:HDd}, that both the number of non-zero terms and  dimensionality of the conventional solution  increases as more samples become available. The reason is that it gets more challenging to meet the desired $\epsilon$ with more samples considering that the measurement matrix lacks some essential bases.

\subsection{Example 4: A community land model} \label{sec:clm}
Previous examples were designed such that only a small subset of random inputs have a significant impact on output. In this example, we consider a real world problem were the number of significant random inputs is not \emph{a priori} known. We aim to develop a PC surrogate for a computationally expensive community land model (CLM) using our incremental algorithm. CLM is the land component of community earth system model (CESM). CLM is used to predict the future states of climate and ecosystems including the future condition of carbon cycles and vegetation dynamics. In this paper, we consider the community land model with carbon/nitrogen biogeochemistry (CLM-CN). In \cite{sargsyan2014dimensionality}, offline CLM-CN simulations were performed to predict the steady-state vegetation for Niwot Ridge flux tower site in Colorado for a 100 year time horizon. Five different output quantities of interest were considered: leaf area index, total vegetation carbon, gross primary production, heterotrophic respiration and photosynthesis. These models are computationally expensive, with the computation time of a single simulation reaching 10 hours on a single processor \cite{sargsyan2014dimensionality}. Thus, a surrogate for this expensive model can significantly reduce the simulation cost and allow further analysis. In this paper, we seek to develop a surrogate model for the calculation of leaf area index ($u_{LAI}$), as a scalar QoI, using a PC expansion over the input parameters. Out of the 81 uniformly distributed CLM input parameters  listed in \cite{sargsyan2014dimensionality}, 2 parameters can be written as functions of the other parameters. Therefore, only 79 parameters are independent and are used as random inputs in the Legendre-based PC expansion. 

In \cite{sargsyan2014dimensionality}, when the QoI, $u_{LAI}$, is smaller than 0.3, it was assumed to that the system is in a dead vegetation state. The probability space, $\Omega$, is then considered to consist of two mutually exclusive subspaces; one that results in dead vegetation state, and one with $u_{LAI} \ge 0.3$, which is denoted by $\mathcal{A}$. The reason that $u_{LAI}\leqslant  0.3$ is considered as dead vegetation is that it is believed that the simulations resulting in small $u_{LAI}$ would have converged to zero if simulations were run for a longer time. In this work, we seek to approximate the output over the latter subspace, by  building a PCE for the live vegetation state. We do so by only  using the samples with $u_{LAI} \ge 0.3$. Specifically, the sample set includes 9983 simulation results, out of which 3437 samples belong to live vegetation state  \cite{sargsyan2014dimensionality}.

Since $u_{LAI}$  is by definition takes non-negative values, the logarithm of $u_{LAI}$  is approximated with a PCE, given by
\[
 u_{LAI} (\bm \Xi)= 
\begin{cases}
    \text{exp}( \sum_{\bm \alpha} c_{\bm \alpha } \psi_{\bm \alpha}(\bm \Xi)), & \text{if} \quad \bm \Xi \in \mathcal{A},\\
    0,              & \text{otherwise}.
\end{cases}
\]

We estimate the PC coefficients using both conventional and incremental algorithms. For conventional approach, we include all 79 random inputs and evaluate the measurement matrix for $k=2$. In this example, instead of prefixing the inaccuracy tolerance $\epsilon$, we use cross-validation approach to select $\epsilon$, to avoid overfitting. Cross-validation results showed that $\epsilon=0.3 \left \| \bm u \right \|_{2}$ results in the lowest prediction errors computed using validation samples. Although this accuracy level might seem too low, it is in fact reasonable considering the high dimensionality of the problem and the limited number of samples. 

\begin{figure}
	\begin{subfigure}[t]{0.49 \linewidth}
		\includegraphics[width=.95\linewidth]{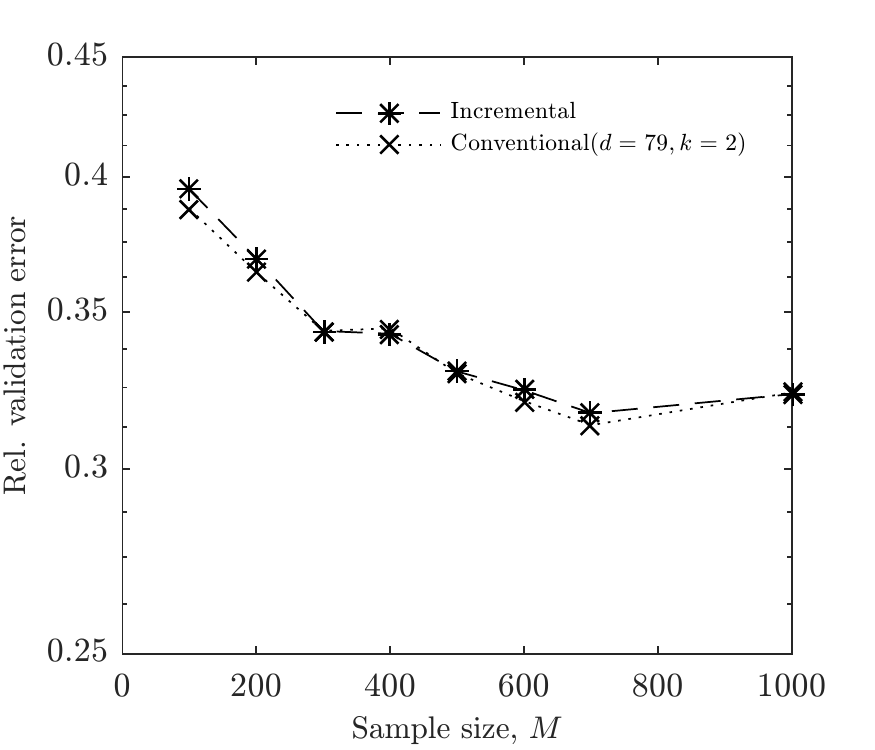}
		\caption{}
		\label{fig:clma}
	\end{subfigure}
	\quad
	\begin{subfigure}[t]{0.49\linewidth}
		\includegraphics[width=.95\linewidth]{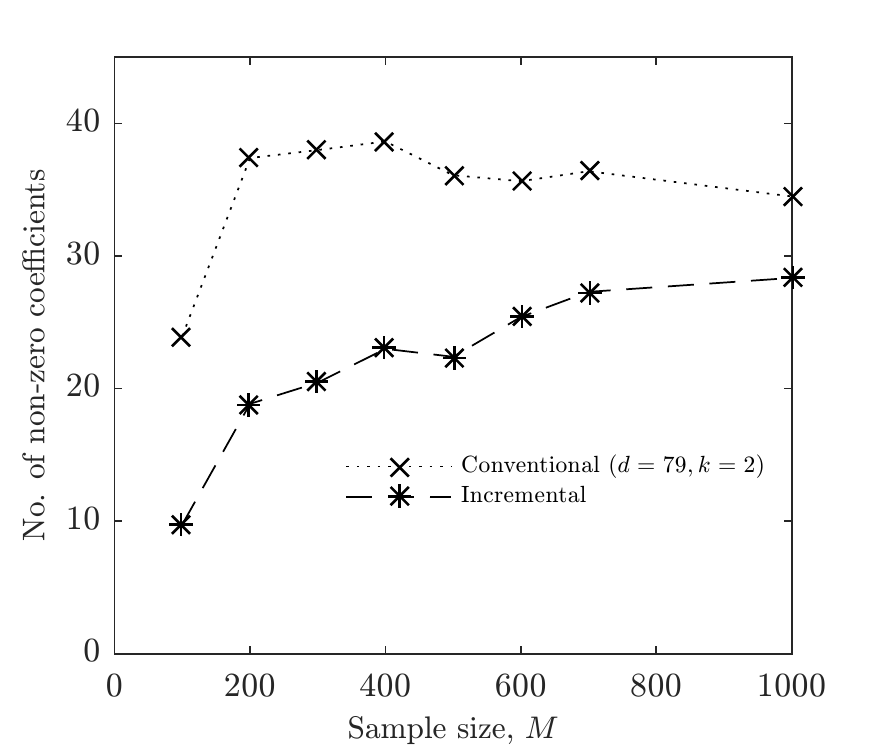}
		\caption{}	
		\label{fig:clmb}
	\end{subfigure}
	\\
	\begin{subfigure}[t]{\linewidth}
		\centering
		\includegraphics[width=0.49 \linewidth]{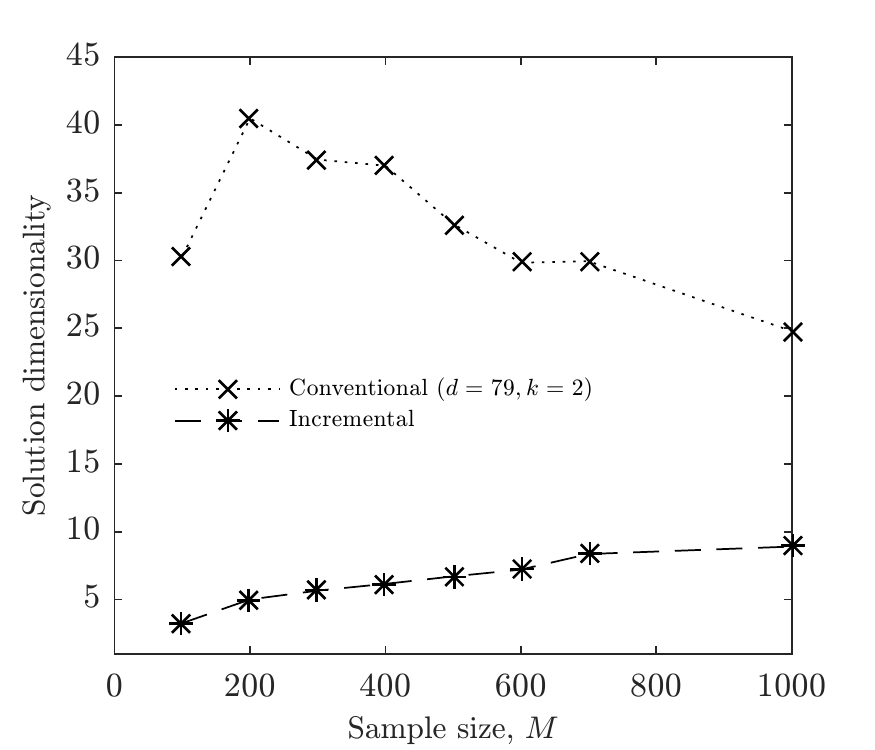}
		\caption{} 
		\label{fig:clmc}
	\end{subfigure}
		\captionsetup{}
		\caption{Comparison of incremental algorithm and conventional BPDN results for community land model (a) relative validation error (b) number of non-zero coefficients (c) solution dimensionality.} 
		\label{fig.example3}
	\end{figure}

Figure \ref{fig.example3} shows the average results for 20 independent trials for incremental and conventional algorithms. Figure \ref{fig:clma} compares the average relative validation error for 200 test samples. Although incremental algorithm does not result in significant reduction in prediction error, i.e., most probably because of very small sample sizes, as it can be seen in Figure \ref{fig:clmb} and \ref{fig:clmc}, it considerably reduces the dimensionality and sparsity of the approximation solution.  Invoking Occam's razor principle, such lower dimensional models may be preferred over  higher dimensional ones~\cite{blumer1990occam}.

\begin{figure}[H]
  	\centering
 	\includegraphics[width=.5\linewidth]{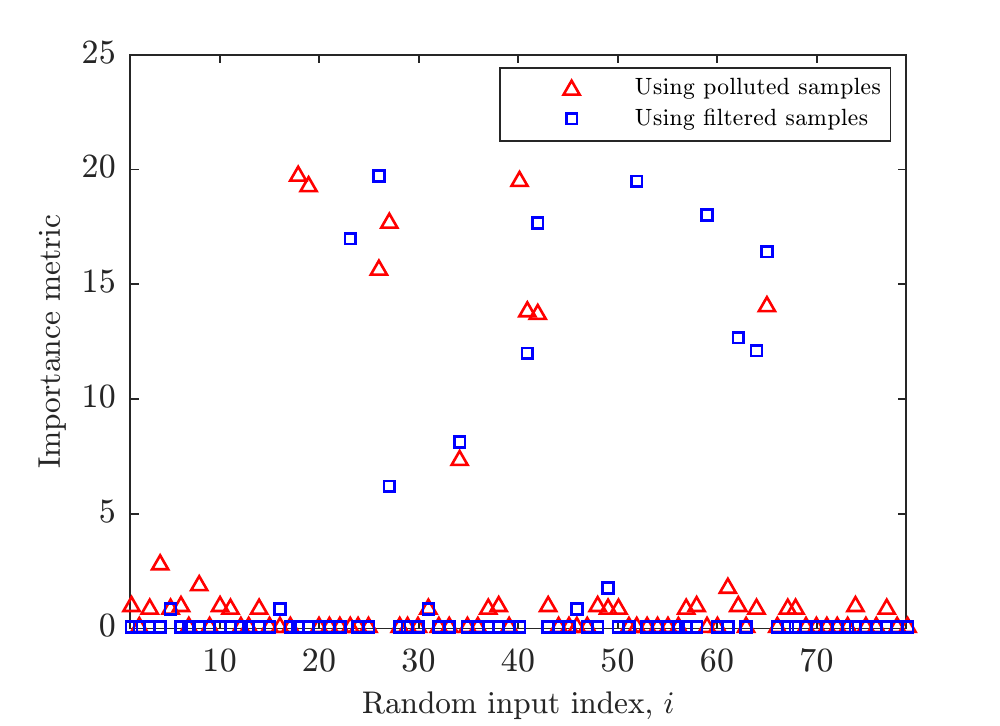}

 \captionsetup{}
 \caption{Importance metrics calculated for all random inputs according to Eq.~\ref{importance matrix}. Filtered samples include only the realizations that lead to non-dead vegetation state. Polluted samples consist of filtered samples (85\%) and realizations leading to a dead vegetation state (15\%).} 
 \label{fig.importance}
\end{figure}

To evaluate how successful our algorithm is in dimension reduction and identifying the right significant parameters, we compare our results with the variance-based sensitivity analysis performed in \cite{sargsyan2014dimensionality} using the PC surrogate estimated by their proposed Bayesian compressive sensing (BCS).  To do so, we first define an importance metric quantified for each random input and averaged over independent trials. A number of independent trials are considered merely in order to make the results less sensitive to a particular sample set, as the problem is severely undersampled. Our choice of  importance metric considers the number of times an input parameter, i.e. a particular dimension, appears in the identified sparsest solution across all trials. It also notes the order in which parameters are added to optimal dimension during the incremental algorithm, i.e. its order in the final dimension reduced input vector $\bm \Xi^r$. Specifically, for each random input, $\Xi_{i}$, the overall importance metric, $I_{i}$, is defined as
\begin{equation} \label{importance matrix}
I_{i} := \sum_{j=1}^{N_{\text{trials}}} \frac{(d-a_{ij})\delta_{ij}}{d}.
\end{equation}
If dimension $i$ is identified in the sparsest solution in trial $j$, then $a_{i,j}$ denotes its order in the identified input vector,   and $\delta_{ij}=1$; otherwise, $a_{i,j}=0$, $\delta_{ij}=0$. In Figure \ref{fig.importance}, square markers show the importance metrics quantified for all random inputs based on trials for the largest sample sizes, $M=1000$, i.e, $N_{\text{trials}}=20$. We consider a  parameter to be uninfluential if it  appears in the solution of incremental algorithm fewer than 3 times  in the 20 trails.  It can be seen that there is a clear separation between  importance metrics for eleven of input parameters and that for the remaining 68 uninfluential parameters.   

The eleven influential random inputs are shown on the first row of Table~\ref{importance-ranking}. For comparison, the ranking by the variance-based sensitivity/BCS approach reported in \cite{sargsyan2014dimensionality} is also included in the second row of Table~\ref{importance-ranking}.  It seems that our algorithm does not identify the top three influential parameters that are reported in \cite{sargsyan2014dimensionality}, i.e. leaf\_cn, rc\_npool and flnr.  This discrepancy can be explained by noting that in the BCS approach,  realizations that lead to the dead vegetation state are  classified and then filtered out based on an algorithm with 15\% likelihood of misclassification. In order to make a consistent comparison between the two rankings, we simulate their filtering process by  keeping 15\% of realizations with $u_{LAI}  \leqslant  0.3$,  and removing 15\% of realizations with $u_{LAI}  >  0.3$ to form a ``polluted" dataset, out of which $M$ random samples are drawn. We still considered 20 trials with $M=1000$. Based on the newly calculated importance metric, only nine parameters are found to have a large importance metric. The third row in  Table~\ref{importance-ranking} shows the ranking of these nine random inputs. We observe that the nine influential random inputs obtained by our algorithm are the same as nine most influential random inputs recognized in \cite{sargsyan2014dimensionality}, however in different order. As an alternative approach, the parameters are also  ranked by analytically calculating, for the PCE identified by our incremental algorithm, the Sobol' indices  according to \cite{sobol2001global,sudret2008global}. This ranking is shown in the last row of Table~\ref{importance-ranking}, and as it can be seen, is the same as the importance metric $I$ ranking , except for the order of three parameters that are slightly different. In fact, it was observed that the significance of these three parameters are almost equal, both based on their respective Sobol' indices and also based on the importance metric shown in Figure \ref{fig.importance} (the three parameters are indexed at 65, 41, and 42. We conclude that our algorithm is capable of identifying influential random inputs and can be used for dimension reduction. 
\begin{table}[H]
	\centering
 \captionsetup{}
	\caption{Comparison between the rankings of influential parameters in the community land model identified by the Bayesian compressive sampling approach proposed in \protect\cite{sargsyan2014dimensionality}  and the incremental $\ell_1$ minimization (using filtered and polluted samples). The parameters in the shaded cells are identified as uninfluential.}
	\label{importance-ranking}
\scalebox{0.65}{
	\begin{tabular}{|c|c|c|c|c|c|c|c|c|c|c|c|}
		\hline
		\textbf{Algorithm} & \multicolumn{11}{c|}{\textbf{Ranking of random inputs (left to right)}}                                  \\ \hline
		\textbf{\begin{tabular}[c]{@{}c@{}}Incremental/metric $I$\\ filtered data \end{tabular}} & {foot\_leaf} & rf\_s3s4  & k\_s4    & q10\_mr   & frootcn & r\_mort   & dnp   & q10\_hr & br\_mr   & leaf\_long & stem\_leaf \\ \hline
		\textbf{\begin{tabular}[c]{@{}c@{}}BCS approach\\ of \cite{sargsyan2014dimensionality}\end{tabular}}              & leafcn    & rc\_npool & froot\_leaf & leaf\_long & flnr   & r\_mort & stem\_leaf & br\_mr     & q10\_mr &  rf\_s3s4   &  dnp   \\ \hline
		\textbf{\begin{tabular}[c]{@{}c@{}}Incremental/metric $I$\\ polluted data\end{tabular}} & leafcn & rc\_npool & flnr       & stem\_leaf & froot\_leaf & r\_mort  & br\_mr & q10\_mr & {leaf\_long} & \cellcolor{blue!25} qe25 & \cellcolor{blue!25} rhosnir\\ \hline
		\textbf{\begin{tabular}[c]{@{}c@{}}Incremental/Sobol'\\ polluted data\end{tabular}} & leafcn & rc\_npool & flnr       & stem\_leaf & froot\_leaf & br\_mr  & q10\_mr & r\_mort & {leaf\_long} & \cellcolor{blue!25} qe25 & \cellcolor{blue!25} rhosnir \\ \hline
	\end{tabular}}

\end{table}

\subsection{Other penalty functions}
The proposed incremental approach is not limited to $\ell_{1}$ minimization problems, and can improve the performance of other variates of sparse recovery algorithms that use  penalty functions other than $\ell_1$-norm, such as weighted $\ell_1$ minimization~\cite{candes2008enhancing}, $\ell_p$ minimization~\cite{xu2012regularization}, and $\ell_1-\ell_2$ minimization algorithms~\cite{yin2015minimization}. The reason is that if the PC measurement matrix does not contain the bases in which the PCE is sparsest in, regardless of what penalty function is used, the solution's accuracy is heavily impaired. Also, in all compressive sample approaches, a highly coherent measurement matrix always leads to a lower solution sparsity and accuracy, regardless of the type of sparsity promoting penalty function. To demonstrate this in an example,  let us consider the $\ell_{1}-\ell_{2}$ minimization algorithm. Numerical results have shown that using $\ell_{1}-\ell_{2}$-norm instead of the widely used $\ell_{1}$-norm can lead to better performance \cite{yin2015minimization}. Theorem \ref{Theory3} establishes bounds on $\ell_{1}-\ell_{2}$ recoverability. 

\begin{theorem}[\cite{yin2015minimization}] \label{Theory3}
Let $\bm c$ be an S-sparse vector satisfying 
	\begin{equation}
	a(S):= \left ( \frac{\sqrt{3S}-1}{\sqrt{S}+1} \right )^{2} >1.
	\end{equation}
Also, let $\bm u= \bm \Psi \bm c ^*+ \bm z$ with $\left \| \bm z \right \|_2 < \epsilon$, and suppose $\bm \Psi$ satisfies the condition
	\begin{equation}
	\delta_{3S}+ a(S)\delta_{4S} < a(S)-1.
	\end{equation}
Then $\overline{\bm c}$, the solution of $\ell_{1}-\ell_{2}$ minimization problem
	\begin{equation} \label{eq:consl1l2min}
	\underset{\bm c}{\textrm{min}}\left \| \bm c \right \|_{1}-\left \| \bm c \right \|_{2} \quad \textrm{subject to} \quad \left \| \bm u-\bm \Psi \bm c \right \|_{2}\leqslant \epsilon,
	\end{equation}
	satisfies 
	\begin{equation}
	\left \| \bm c^{*}-\overline{\bm c}\right \|_{2} \leqslant \frac{2 \epsilon \sqrt{1+a(S)}}{\sqrt{a(S)(1-\delta_{4S})}-\sqrt{1+\delta_{3S}}}.
	\end{equation}
\end{theorem}
According Proposition \ref{prop1}, smaller coherence for $\bm \Psi$	enforces smaller $\delta_{3S}$ and $\delta_{4S}$ and results in smaller bound on the recovery error, and vice versa. Therefore, the issue of coherence is present in $\ell_{1}-\ell_{2}$ minimization algorithm, as well, and as such our approach is expected to improve the results. 

\begin{figure}[H]

	\begin{subfigure}[t]{0.49 \linewidth}
		\centering
		\includegraphics[width=.95\linewidth]{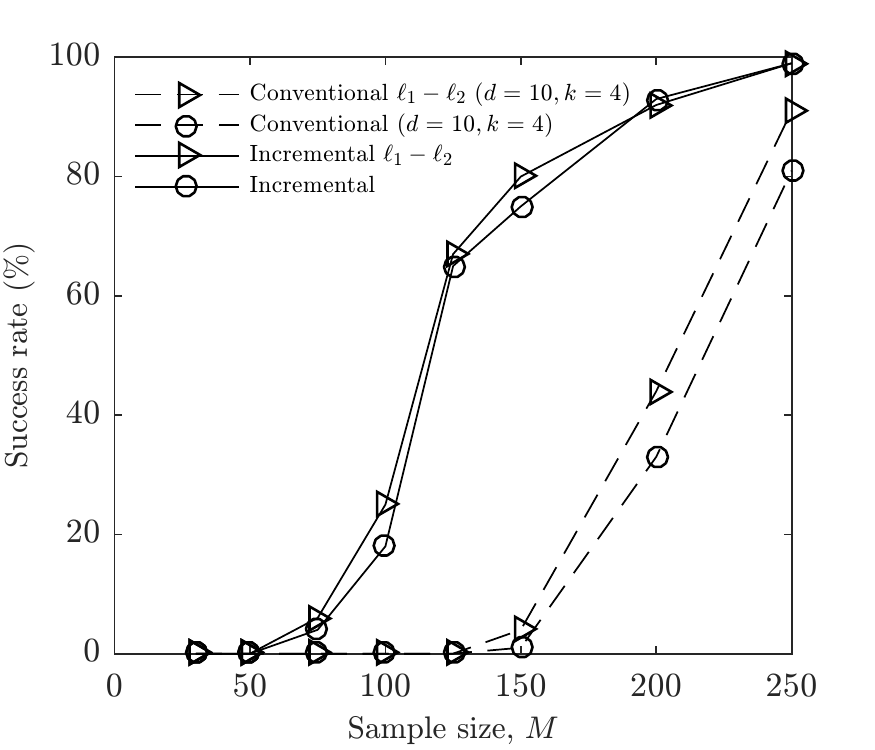}
		\caption{}
		\label{fig:L1L2a}		
	\end{subfigure}
	\quad
	\begin{subfigure}[t]{0.49 \linewidth}
		\centering
		\includegraphics[width=.95\linewidth]{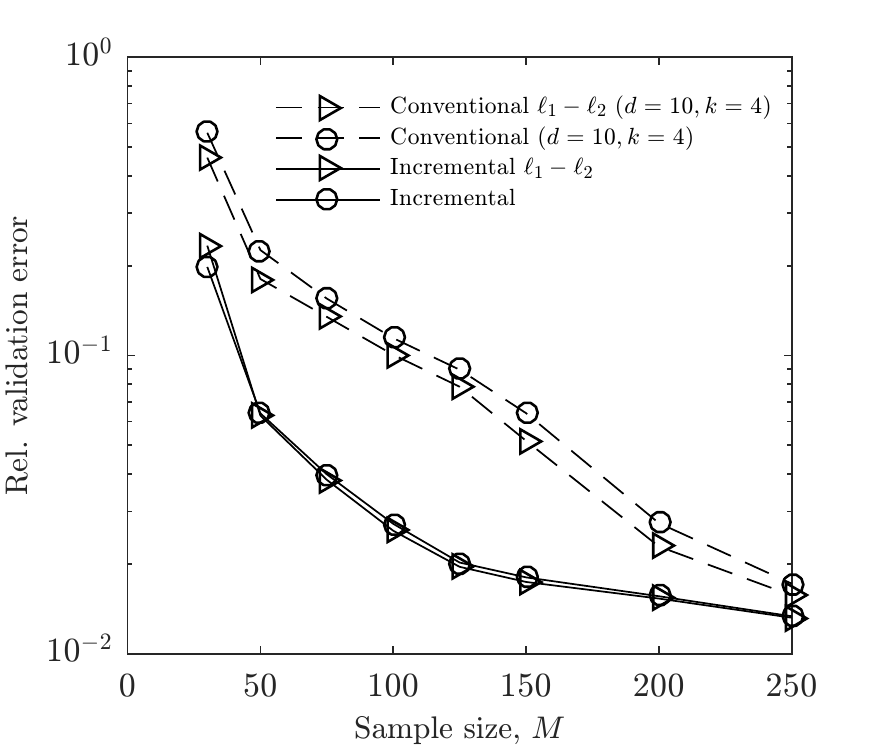}
		\caption{}
		\label{fig:L1L2b}
	\end{subfigure}

	\captionsetup{}
	\caption{Comparison of incremental and conventional $\ell_{1}-\ell_{2}$ minimization results for manufactured polynomial expansion (a) success rate (b) relative validation error.} 
	\label{fig.L1L2}
\end{figure}

To numerically demonstrate this, let us consider the target function used in Section~\ref{sec:manufacPCE10d}, and the sparse recovery of the associated PC approximations using $\ell_1$ and $\ell_{1}-\ell_{2}$ minimization algorithms. We make use of the algorithm developed in \cite{yin2015minimization} for $\ell_{1}-\ell_{2}$ minimization, which solves the following unconstrained minimization 
\begin{equation} \label{eq:unconsl1-2min}
\underset{\bm c}{\textrm{min}} \left \| \bm u-\bm \Psi \bm c \right \|_{2}^{2}+ \lambda(\left \| \bm c \right \|_{1}-\left \| \bm c \right \|_{2}).
\end{equation}
To make this unconstrained minimization equivalent to the constrained minimization given in (\ref{eq:consl1l2min}), we used an iterative approach to calculate the corresponding $\lambda$ that matches the prescribed $\epsilon$. Figure \ref{fig.L1L2} compares the results for conventional and incremental $\ell_{1}-\ell_{2}$ minimization. It can be seen that regardless of the penalty function, incremental algorithm significantly         outperforms the conventional algorithm.
 
\section{Conclusion}
In this paper, we proposed an incremental compressive sampling algorithm for sparse approximation of PC expansions, when limited number of samples are available. The proposed algorithm seeks to identify the sparsest PC approximation, among expansions with various  dimensions and orders. We demonstrated the effectiveness of these algorithms in problems where the solution lies in a lower dimensional subspace in the parameter space. The proposed approach resolves the challenge posed by highly coherent measurement matrices, relevant in high dimensional expansions, by first constituting  a very low dimensional candidate solution, and then incrementally adding its dimension or order, depending on the improvement in sparsity.   Numerical examples are provided to validate the proposed incremental algorithm. The results show that a substantial improvement in solution sparsity and accuracy can be achieved by using the incremental algorithm instead of conventional compressive sampling approach.

  Our algorithm is different from  the adaptive algorithm proposed in~ \cite{jakeman2015enhancing}, in terms of the criterion for selecting new bases or stopping the algorithm, and also the choice of initial basis set.  Another important distinction is that if the user has a specific preference for the inaccuracy  tolerance $\epsilon$, our algorithm will provide the associated sparsest solution. In other words, our incremental algorithm can perform a tolerance-dependent sparse recovery or dimensionality reduction. Specific tolerance levels can, for instance, be rigidly prescribed by a user based on expert judgment or  knowledge about the measurement noise, instead of a cross-validation procedure.   

The proposed  algorithm can be further refined to become more efficient by an anisotropic trial basis expansion. Specifically, one can refine how expansion order is increased in each trial in the algorithm. Currently, at a dimensionality level, a trial order increase is done by incrementing the total order by one and isotropically impacting all the dimensions. A more refined anisotropic basis set expansion can involve at each iteration multiple one-dimensional order increment trials, each increasing the order in one of the dimensions by one. This can prevent undesirably large increases in the coherence after a one-step degree increment in the algorithm. For problems with a small underlying dimensionality, this gain may not be substantial and the current form of the algorithm is expected to perform well. In cases with larger underlying dimensionality, however, such refinement can potentially lead to a  more precise  identification of the sparsest solution.

\section*{Acknowledgement}
The authors would like to acknowledge Dr. Khachik Sargsyan for providing the  Community Land Model simulation data which was used in the numerical example of Section~\ref{sec:clm}, and is described in detail in~\cite{sargsyan2014dimensionality}.     

\section*{References}

\bibliography{bibfile}
%\bibliography{full_ref}

\end{document}